\documentclass{aa}
\usepackage{psfig}
\def\ha{H$\alpha$}
\def\h2{H{\small\,II}}

\newcommand{\msun}{$M_\odot$}

\newcommand{\etal}{et al.}
\newcommand{\hb}{\ifmmode {\rm H}\beta \else H$\beta$\fi}
\newcommand{\whb}{\ifmmode EW({\rm H}\beta) \else $EW({\rm H}\beta)$\fi}
\newcommand{\wha}{\ifmmode EW({\rm H}\alpha) \else $EW({\rm H}\alpha)$\fi}
\newcommand{\hii}{H~{\sc ii}}
\newcommand{\hei}{He~{\sc i}}
\newcommand{\niiib}{N~{\sc iii} $\lambda$4512}
\newcommand{\nv}{N~{\sc v} $\lambda$4612}
\newcommand{\niii}{N~{\sc iii} $\lambda$4640}
\newcommand{\civb}{C~{\sc iv} $\lambda$4658}
\newcommand{\heii}{He~{\sc ii} $\lambda$4686}
\newcommand{\ciii}{C~{\sc iii} $\lambda$5696}
\newcommand{\civ}{C~{\sc iv} $\lambda$5808}
\newcommand{\oh}{12 + $\log$(O/H)}
\newcommand{\mup}{\ifmmode M_{\rm up} \else $M_{\rm up}$\fi}
\newcommand{\dt}{\ifmmode \Delta t \else $\Delta t$\fi}
\def\aap{A\&A}

\def\aaps{A\&AS}
\def\aj{AJ}
\def\apj{ApJ}
\def\apjl{ApJ}
\def\apjs{ApJS}
\def\mnras{MNRAS}

\begin{document}

  \thesaurus{  03          
              (11.01.1;    
               11.05.2;    
               11.09.4;    
               11.19.3;    
               11.19.5;     
               08.12.3;    
               08.23.2  )} 

   \title{Massive star populations and the IMF in metal-rich starbursts
     \thanks{Based on observations at Kitt Peak National Observatory, National
       Optical Astronomical Observatories, operated by the Association of
       Universities for Research in Astronomy, Inc., under contract with the
       Nationsal Science Foundation}}

   \author{Daniel Schaerer 
    \inst{1}
    \and
    Natalia G. Guseva \inst{2}
    \and
    Yuri I. Izotov \inst{2}
    \and
    Trinh X. Thuan \inst{3}}
   \offprints{D. Schaerer, schaerer@obs-mip.fr}
   \institute{   Laboratoire d'Astrophysique, 
                 Observatoire Midi-Pyrenees, 14, Av. E. Belin,
                 F-31400 Toulouse, France
              \and
                 Main Astronomical Observatory
                 of National Academy of Sciences of Ukraine,
                 Goloseevo, 03680, Kiev-127, Ukraine
              \and
                 Astronomy Department, University of Virginia,
                 Charlottesville, VA 22903, USA}
\date{Received 2 march 2000 / Accepted 31 august 2000}
\maketitle


\begin{abstract}
We present new spectroscopic observations of Mkn 309, a starburst galaxy 
with one of the largest WR populations known. A highly super solar metallicity 
of \oh\ $\sim$ 9.3--9.4 is derived.
Using additional objects from Guseva \etal\ (2000) we analyse a sample of five 
metal-rich ([O/H] $>$ 0) WR galaxies with the main goal of constraining
 the basic 
properties of the massive star populations (IMF slope, \mup) and the star formation 
history (age, burst duration) of these objects by quantitative comparisons 
with evolutionary synthesis models. 
The following main results are obtained:

\begin{itemize}
\item The observations are well explained by extended bursts of star formation with 
durations $\Delta t \sim$ 4--10 Myr seen at ages of 7--15 Myr
or a superposition of several bursts with age differences of $\sim$ 4--10 Myr
including a young ($\la$ 5 Myr) burst.
This naturally explains both the observed WR populations (including WN and
WC stars) and the presence of red supergiants. 
The burst durations, somewhat longer compared to those derived in other WR 
galaxies using the same models (Schaerer \etal\ 1999a), 
are plausible in view of the physical sizes of the observed regions and the 
nature and morphology of our objects (nuclear starbursts), and pose no fundamental 
physical problem.

\item The SEDs in the optical range are very well reproduced for all 
objects, provided the stellar light suffers from a smaller extinction than
that of the gas (derived from the Balmer decrement).
This confirms earlier findings from studies combining UV--optical data
of other starburst galaxies.

\item All the considered observational constraints are compatible with a 
Salpeter IMF extending to masses \mup\ $\ga$ 40 \msun.
Adopting a conservative approach we derive a {\em lower limit} of 
\mup $\ga$ 30 \msun\ for the Salpeter IMF.
From more realistic assumptions on the metallicity and SF history we
favour a lower limit \mup $\ga$ 30--40 \msun, which is also in agreement
with \hb\ equivalent width measurements of metal-rich \hii\ regions
in spiral galaxies indicating an upper mass cut-off of at least $\sim$ 
35 -- 50 \msun. 
Steep IMF slopes ($\alpha \ga 3.3$) are very unlikely.

\end{itemize}

The uncertainties of our results are discussed. We compare our findings to
other work on massive star populations and the IMF in similar environments.
We stress the importance of direct analysis of stellar populations
compared to other indirect methods based on properties of ionized gas
to constrain the IMF in metal-rich starbursts.


\end{abstract}

\section{Introduction}
%
Massive galaxies, the central regions of galaxies, AGNs, and related objects
are generally expected to harbour metal-rich gas and stars, as
indicated by numerous observations (e.g.\ Hamann 1997, Henry \& Worthey 1999).

For the understanding of actively star-forming galaxies (also loosely denoted as 
``starbursts'' here) knowledge about massive stars 
--- their evolution, as well as radiative, mechanical, and chemical feedback --- 
and their relative number compared to lower mass stars is of prime importance.
However, the properties and evolution of individual metal-rich high mass stars,
massive star populations, and their initial mass function (IMF) in metal-rich 
environments are still poorly known.

Although of great interest to a number of astrophysical problems
few quantitative studies of {\em individual massive stars} at high metallicities
exist so far. See e.g.\ McCarthy \etal\ (1997) and Monteverde \etal\ (1997) for
work on M31 and M33 stars, Najarro \etal\ (1997) and Figer \etal\ (1999)
for Galactic Center stars, and Maeder \& Meynet (1994) for the use of WR/O star
statistics.
Studies of {\em integrated stellar cluster spectra}, such as the pioneering work of Bica,
Alloin and coworkers (Bica \& Alloin 1986), have provided an extremely useful base
for the understanding of stellar populations in more complex objects.
However, templates of both young and metal-rich clusters (metallicities above
$\sim$ 1.25 $\times$ solar) are missing. 
This clearly shows that studies of massive stars, both individual stars or integrated 
populations, with super-solar metallicity remain largely unexplored territory.
While in principle nebular lines can be used through photoionization modeling
to constrain the stellar content of \hii\ regions (e.g.\ Garc\'{\i}a-Vargas \etal\ 1995,
1997, Stasi\'nska \& Leitherer 1996) the inherent difficulties of optical studies 
at high metallicities (e.g.\ strong sensitivity of lines to unknown electron
temperature) render this approach rather uncertain. 
Here, we preferentially rely on the analysis of direct stellar features.

Of particular interest for studies of metal-rich starbursts is the upper part of the
IMF. More precisely, important questions are what is the slope of the IMF and
what are the most massive stars formed in such environments?
From work undertaken in recent years the picture of a ``universal'' IMF
with a slope close to the Salpeter value for the mass range of interest 
here ($M \ga 5$ \msun) and the existence of massive stars (\mup $\sim$ 60--100
\msun) seems to emerge (e.g.\ Larson 1998, references in Gilmore \& Howell 1998,
but also Scalo 1998).
So far these results are found to hold for a diversity of objects and metallicities
from solar to very metal poor environments (SMC, blue compact 
dwarf galaxies with \oh $\ga$ 7.2). 
Very little is, however, known about the IMF at high metallicities.

Indirect indications for a possible systematic difference of the IMF come
e.g.\ from studies on IR-luminous starburst galaxies, where generally a 
soft radiation field is observed, which, among other explanations, 
can be interpreted as a deficit of massive stars (cf.\ Goldader \etal\ 1997, 
Luhman \etal\ 1998). 
Given the high luminosity and intense star-formation in these objects,
the metallicity is likely high and could be responsible for the IMF change.
Most recently Bresolin \etal\ (1999) claim a low value of the upper mass cut-off
(\mup\ $\sim$ 25 \msun) at super-solar metallicity from considerations of the nebular 
properties of giant extragalactic \hii\ regions.
Although not yet well established (cf.\ e.g.\ the recent modeling of ISO observations
by Thornley \etal\ 2000), these results, if true, have many implications.
It is important, however, to note that the only indications for a low value of \mup\ 
in these objects come from the {\em properties of the gas} which is ionized by 
massive stars. Many causes (dust absorption, nebular geometry, uncertainties in 
stellar atmosphere models etc.) could be responsible for a misinterpretation in 
terms of \mup. A direct probe of the {\em stellar content} of metal-rich objects
is needed. A first step in this direction is done in the present work.

The so-called Wolf-Rayet (WR) galaxies (cf.\ Conti 1991, Schaerer \etal\ 1999b),
where broad emission lines from WR stars observed in the integrated spectra
indicate the presence of massive stars ($M_{\rm initial} \ga$ 25 \msun), 
provide an extremely interesting opportunity to constrain evolutionary models
for massive stars in different environments and to probe the upper part of the IMF.
These issues have first been explored in the important studies of Kunth 
\& Joubert (1985), Arnault \etal\ (1989) and Vacca \& Conti (1992). 
Recent progress comes from new high S/N 
observations (e.g.\ Schaerer \etal\ 1999a -- hereafter SCK99, Guseva \etal\ 2000 
-- hereafter GIT00) and the use of improved evolutionary synthesis
models (Cervi\~no \& Mas-Hesse 1994, Meynet 1995, Schaerer 1996, Schaerer \& Vacca 
1998, hereafter SV98).
Reviews on the issues addressed in studies of WR galaxies 
can be found for example in the volume of van der Hucht \etal\ (1999) and in
Schaerer (1999ab).

The sample of GIT00 includes also some metal-rich WR galaxies. 
In the present work we extend this sample and present a detailed analysis of five 
objects with super solar metallicity. Our main goal is to determine the basic 
properties of the massive star populations (IMF slope, \mup) and the star formation 
history (age, burst duration) in these metal-rich starbursts.

The structure of this paper is as follows:
The observational data is described in Sect.\ 2.
Physical conditions and abundances are derived in  Sect.\ 3.
The stellar content of our objects is analysed in Sect.\ 4.
In Sect.\ 5 we derive constraints on the star formation properties
and massive star populations from comparisons with evolutionary synthesis models.
The interpretation in terms of the IMF is discussed in Sect.\ 6.
Our main results are summarised in Sect.\ 7.

\section{Observational data}

%
\begin{table*}[htb]
\caption{General properties of the metal-rich WR galaxies}
\label{tab_general}
\begin{center}
\begin{tabular}{llllllllll} \hline
Galaxy & $\alpha$(1950) & $\delta$(1950) & $m_B$ & Type      \\ \hline
Mkn 589  & 02 11 08.7 & +03 52 08 & 14.51 & S? \\
Mkn 1199 & 07 20 28.5 & +33 32 24 & 13.7  & Sc \\
Mkn 710  & 09 52 10.2 & +09 30 32 & 13.04 & SB(rs)ab Sbrst \\
Mkn 1259 & 10 36 03.0 & --06 54 37 & 13.5  & S0 pec?  \\
Mkn 309  & 22 50 10.0 & +24 27 52 & 15.44 & Sa Sy2 \\ 
\hline
\end{tabular}
\end{center}
\end{table*}

The observational data used in the present paper consist of
four high metallicity ($\sim$ solar and above) objects taken from the 
large sample of Wolf-Rayet galaxies
of GIT00 (Mkn 589, Mkn 710$\equiv$NGC 3049, Mkn 1199, Mkn 1259) ,
supplemented by the data on NGC 3049 by Schaerer \etal\ (1999a), and new
observations of Mkn 309. 
The general properties of the objects are given in Table \ref{tab_general}.
Distances and derived metallicities are included in Table \ref{Tab5} below.
Except for Mkn 710, a bright \hii\ region in a barred spiral galaxy,
the star forming regions discusses here are fairly compact central regions
of galaxies (``nuclear starbursts'').
Note that this is in contrast with the remaining WR galaxies of GIT00,
comprising mostly metal-poor blue compact dwarf galaxies.


%
\begin{figure*}[htb]
\centerline{\psfig{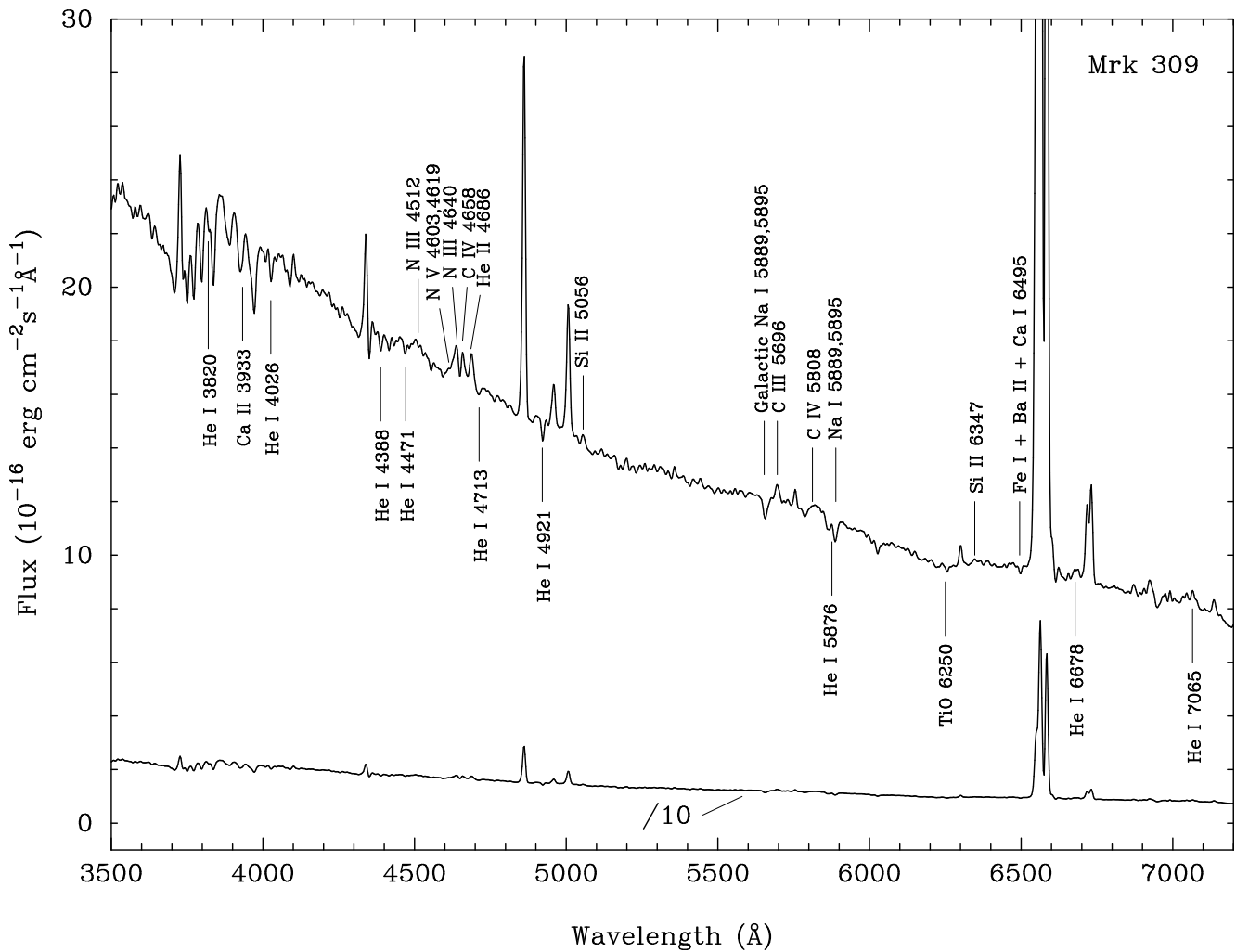}}
\caption[]{\label{Fig1}
Observed spectrum of Mkn 309. Some heavy element absorption features, 
He {\sc i} and WR lines are marked. All He {\sc i}
lines in blue part of the spectrum are in absorption while He {\sc i} $\lambda$6678
and $\lambda$7065 lines are in emission.
}
\end{figure*}

   Spectroscopic observations of Mkn 309 were made with the Kitt Peak 4m 
Mayall telescope during a clear night on 18 June 1999 with the Ritchey-Chr\'{e}tien 
RC2 spectrograph used
in conjunction with the T2KB 2048$\times$2048 CCD detector. We use a 
2\arcsec$\times$300\arcsec\ slit with the 
KPC--10A grating (316 lines mm$^{-1}$) 
in first order, with a GG 375 order separation filter. This filter cuts off 
all second-order contamination for wavelengths blueward of 7200\AA\, which is the 
wavelength region of interest here. The above instrumental set-up gave a 
spatial scale along the slit of 0.69 arcsec pixel$^{-1}$, a scale perpendicular 
to the slit of 2.7\AA\ pixel$^{-1}$, a spectral range of 3500--7300\AA\ and a 
spectral resolution of $\sim$ 6\AA. The seeing was 1\arcsec. 
The total exposure time was 60 minutes and was broken up into 2 subexposures.
The slit was oriented close to the parallactic angle, so that no 
correction for atmospheric differential refraction was necessary.

     The Kitt Peak IRS spectroscopic standard star Feige 34 was observed
for flux calibration with a 6\arcsec\ wide slit. 
Spectra of He--Ne--Ar comparison arcs were obtained
before and after each observation to calibrate the wavelength scale. 

     The two-dimensional spectra were bias subtracted and flat-field corrected.
We then use the IRAF\footnote[2]{IRAF is distributed by National Optical
Astronomical Observatories, which is operated by the Association of Universities 
for Research in Astronomy, Inc., under cooperative agreement with the National 
Science Foundation.} software 
to perform wavelength calibration, correction for distortion and tilt for each 
frame and extraction of one-dimensional spectra.
The extracted spectra from each frame were coadded
and cosmic rays were removed manually. 
The sensitivity curve was obtained from the observed spectral energy
distribution (SED) of Feige 34 with a high-order polynomial.
The intensity distribution of the continuum and of the emission lines
along the slit shows a strong maximum with FWHM = 2\arcsec. 
We extract a one-dimensional spectrum of the brightest part of 
the galaxy in a 2\arcsec$\times$4\farcs2 aperture.
The signal-to-noise ratio for this spectrum is $\sim$ 100.
The spectrum for Mkn 309, uncorrected for extinction, is shown in Figure \ref{Fig1}. 

We detect several Wolf-Rayet emission lines. The He {\sc i} lines in the blue 
part of the spectrum are in absorption suggesting the presence of early B stars 
in the star-forming region. 
We also mark several absorption features of heavy elements, such as Ca {\sc ii} 
$\lambda$3933, Na {\sc i} $\lambda$5889, 5895, the TiO $\gamma^\prime$ band at 
$\lambda$6250 and the Fe {\sc i} + Ba {\sc ii} + Ca {\sc i} 
$\lambda$6495 blend.  Although weak, the latter feature is 
not an artifact. It is seen in the spectra of all studied galaxies (Fig.\ 2) 
because of their very high signal-to-noise ratio.
Although Bica \& Alloin (1986) note that the latter blend
(their window \#59) includes also CN and TiO features, the narrowness of the
blend suggests that contribution of molecular bands is rather small.

The observed line intensities have been
corrected for interstellar extinction using the reddening law by Whitford (1958).
We decided not to correct 
line intensities separately for the Galactic and intrinsic extinctions, partly,
because of the large discrepancy between the determinations of 
$A_B$ = 0.649  by Schlegel et al.\ (1998) and $A_B$ = 0.190 by Burstein \& Heiles 
(1982). Instead we derive extinction
directly from the observed hydrogen line intensities which
have been also corrected for underlying stellar absorption, with 
the equivalent width for absorption hydrogen lines derived self-consistently
together with the extinction coefficient from the observed intensities of all
hydrogen lines. The redshift $z$ = 0.04174$\pm$0.00009 derived from
the observed wavelengths of 16 emission lines is close to $z$ = 0.04215 in
the NED.
For the latter heliocentric redshift, 1\arcsec\ = 817 pc, assuming 
$H_0$ = 75 km s$^{-1}$ Mpc$^{-1}$.  
We show the observed $F$($\lambda$)/$F$(H$\beta$) and 
extinction and absorption-corrected $I$($\lambda$) / $I$(H$\beta$) line 
intensities for Mkn 309 in Table \ref{Tab1}, along with the extinction 
coefficient and the equivalent width of the hydrogen absorption lines, the 
observed flux  and equivalent width of the H$\beta$ emission line. 
The errors of the line intensities include the errors in the placement
of the continuum and the errors introduced by Gaussian fitting of the line
profiles. The indicated uncertainties represent {\em formal} errors.
Note the small intensities of the [O {\sc i}] and [S {\sc ii}] emission
lines suggesting that Mkn 309 is a starburst galaxy, contrary to its 
classification in the NED as Sy2 galaxy.
%
\begin{table}
\caption{Emission line intensities in Mkn 309}\label{Tab1}
\begin{tabular}{lcc} \hline
Ion   &$F$($\lambda$)/$F$(H$\beta$)
&$I$($\lambda$)/$I$(H$\beta$)
 \\ \hline
 3727\ [O {\sc ii}]        &0.40$\pm$0.02& 0.61$\pm$0.03 \\
 4101\ H$\delta$     &0.10$\pm$0.02& 0.25$\pm$0.05 \\
 4340\ H$\gamma$     &0.31$\pm$0.02& 0.45$\pm$0.03 \\
 4363\ [O {\sc iii}]       &0.03$\pm$0.01& 0.03$\pm$0.01 \\
 4861\ H$\beta$      &1.00$\pm$0.02& 1.00$\pm$0.03\\
 4959\ [O {\sc iii}]       &0.15$\pm$0.02& 0.13$\pm$0.02 \\
 5007\ [O {\sc iii}]       &0.38$\pm$0.02& 0.34$\pm$0.02 \\
 5056\ Si {\sc ii}         &0.05$\pm$0.02& 0.04$\pm$0.02 \\
 5755\ [N {\sc ii}]        &0.05$\pm$0.01& 0.03$\pm$0.01 \\
 5876\ He {\sc i}          &0.02$\pm$0.01& 0.02$\pm$0.01 \\
 6300\ [O {\sc i}]         &0.05$\pm$0.01& 0.03$\pm$0.01 \\
 6563\ H$\alpha$     &5.41$\pm$0.10& 3.04$\pm$0.06 \\
 6583\ [N {\sc ii}]        &3.83$\pm$0.07& 2.12$\pm$0.04 \\
 6678\ He {\sc i}          &0.02$\pm$0.01& 0.01$\pm$0.01 \\
 6717\ [S {\sc ii}]        &0.21$\pm$0.01& 0.11$\pm$0.01 \\
 6731\ [S {\sc ii}]        &0.27$\pm$0.01& 0.15$\pm$0.01 \\
 7065\ He {\sc i}          &0.02$\pm$0.01& 0.01$\pm$0.01 \\
 7135\ [Ar {\sc iii}]      &0.05$\pm$0.01& 0.02$\pm$0.01 \\ \\
 $C$(H$\beta$) dex    &\multicolumn {2}{c}{0.68$\pm$0.02} \\
 $F$(H$\beta$)$^a$ &\multicolumn {2}{c}{ 1.74$\pm$0.11} \\
 $EW$(H$\beta$)\ \AA &\multicolumn {2}{c}{11$\pm$1} \\
 $EW$(abs)\ \AA      &\multicolumn {2}{c}{0.8$\pm$0.1} \\ \hline 
\end{tabular}

$^a$in units of 10$^{-14}$ ergs\ s$^{-1}$cm$^{-2}$.
\end{table}

\section{Physical conditions and heavy element abundances in the H II region}
\label{s_abund}

From the observed spectrum we shall now derive the physical conditions
and metal abundances in Mkn 309. For the remaining objects the reader
is referred to GIT00.
Given the weakness or absence of [O~{\sc iii}] $\lambda$4363 in Mkn 309 and
other objects included later for comparison, the abundance determinations
are affected by relatively large uncertainties. These are briefly discussed
below.

\subsection{Mkn 309}
   To determine heavy element abundances, we adopt,
following Izotov et al. (1994, 1997), a two-zone photoionized H II
region model: a high-ionization zone with temperature $T_e$(O {\sc iii}), where
O {\sc iii} and Ne {\sc iii} lines originate, and a low-ionization zone with
temperature $T_e$(O {\sc ii}), where O {\sc ii}, N {\sc ii} and S {\sc ii} lines originate. 
As for the Ar {\sc iii} lines, they originate  in the intermediate zone
between the high and low-ionization regions (Garnett 1992).

   A weak auroral [O {\sc iii}] $\lambda$4363 emission line is detected
with the peak intensity above the 3$\sigma$ level.
However, the intensity of this line is rather uncertain due
to underlying H$\gamma$ stellar absorption and uncertainties in the placement 
of the continuum because of numerous absorption features.
    The temperature $T_e$(O {\sc iii}) calculated using the 
[O {\sc iii}] $\lambda$4363/($\lambda$4959+$\lambda$5007) ratio is unreasonably high 
$\sim$ 47000 K. 
The electron temperature derived from the [N {\sc ii}] $\lambda$5755/$\lambda$6583
ratio $T_e$(N {\sc ii}) = 9900 K, though lower, is still significantly larger than
the electron temperature $T_e$(O {\sc iii}) derived from the empirical relation 
by Edmunds \& Pagel (1984) and can also be subject to the
enhancement mechanisms discussed above.
A similar $T_e$(N {\sc ii}) has been derived by Osterbrock
\& Cohen (1982).
Adopting $T_e$(N {\sc ii}) = 9900K
results in an oxygen abundance 12 + log O/H = 7.85. However, the N/O abundance
ratio in this case is 30 -- 40 times larger than the value expected for
such a low oxygen abundance, which seems unreasonable.
Therefore, we do not use the [O {\sc iii}] $\lambda$4363 and 
[N {\sc ii}] $\lambda$5755 auroral lines for the determination of electron 
temperature. Instead, the empirical method by Edmunds \& Pagel (1984) is used
to derive $T_e$(O {\sc iii}) from the intensities of nebular [O {\sc ii}] $\lambda$3727 and
[O {\sc iii}] $\lambda$4959 + $\lambda$5007 lines. 
The electron temperature for
[O {\sc ii}] is derived using the results of the 
H II photoionization models of Stasi\'nska (1990) and the electron
temperature for [Ar {\sc iii}] is derived using the prescriptions
by Garnett (1992).
The electron number density is calculated from the [S {\sc ii}] 
$\lambda$6717/$\lambda$6731 line intensity ratio. 
Then $T_e$ and $N_e$ are used to derive element abundances.

   In Table \ref{Tab2} we show the derived electron temperatures $T_e$ for different
ions and the electron number density $N_e$. 
The electron temperatures are particularly low, much lower than those derived
from the [O {\sc iii}] and [N {\sc ii}] line intensity ratios. 
We find $N_e$ = 1110 cm$^{-3}$.
The resulting abundances for heavy elements are shown in Table \ref{Tab2}. The oxygen
abundance 12 + log O/H = 9.41 is $\sim$ 3 times larger than the solar value.
In Table \ref{Tab2} we also show the oxygen abundance derived from empirical methods
using [O {\sc ii}] + [O {\sc iii}] intensities (Edmunds \& Pagel 1984) and the [N {\sc ii}] 
$\lambda$6583/H$\alpha$ line intensity ratio (van Zee et al. 1998). All
methods give nearly  the same value of oxygen abundance. 
The nitrogen-to-oxygen abundance ratio is $\sim$ 4 times larger than the solar
value and
within the errors is consistent with the secondary nitrogen production expected
at the high metallicity of Mkn 309.

\begin{table}
\caption{Heavy element abundances in Mkn 309}
\label{Tab2}
\begin{center}
\begin{tabular}{lc} \hline
Parameter&Value \\ \hline
$T_e$(O {\sc iii})(K)                     &3830$\pm$520      \\
$T_e$(O {\sc ii})(K)                      &6110$\pm$520      \\
$T_e$(Ar {\sc iii})(K)                    &4970$\pm$520      \\
$N_e$(S {\sc ii})(cm$^{-3}$)              & 1110$\pm$350      \\ \\
O$^+$/H$^+$($\times$10$^3$)         &0.38$\pm$0.21    \\
O$^{++}$/H$^+$($\times$10$^3$)      &2.20$\pm$2.29    \\
O/H($\times$10$^3$)                 &2.58$\pm$2.30    \\
12 + log(O/H)                       &9.41$\pm$0.39    \\
12 + log(O/H) (Edmunds \& Pagel)    &9.32$\pm$0.20 \\
12 + log(O/H) (van Zee et al.)      &9.32$\pm$0.20 \\ \\
N$^+$/H$^+$($\times$10$^3$)         &0.20$\pm$0.10    \\
ICF(N)                              &6.85\,~~~~~~~~~~   \\ 
log(N/O)                            &--0.29$\pm$0.64~~\\ \\
Ar$^{++}$/H$^+$($\times$10$^6$)     &1.68$\pm$0.75    \\
ICF(Ar)                             &2.25\,~~~~~~~~~~   \\
log(Ar/O)                           &--2.84$\pm$0.41~~\\ \hline
\end{tabular}
\end{center}
\end{table}

\begin{figure}[htb]
\centerline{\psfig{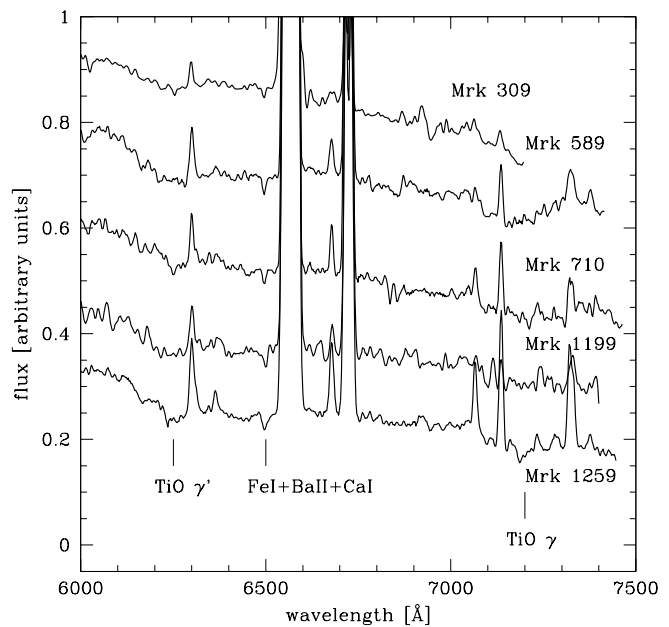}}
\caption{\label{Fig2}
Spectral region showing TiO bands. From top to bottom: Mkn 309, 589, 710, 1199, 1259. 
The narrow Fe {\sc i} + Ba {\sc ii} + Ca {\sc i} $\lambda$6495 blend is also marked.
}
\end{figure}

%
\begin{table}
\caption{Observed parameters of WR lines}
\label{Tab3}
\begin{center}
\begin{tabular}{lc} \hline
Parameter&Value$^a$ \\ \hline
$F$(H$\beta$)$^b$            & 2.20$\pm$0.17 \\
$F$(N {\sc iii} $\lambda$4512) & 0.25$\pm$0.03 \\
$F$(Si {\sc iii} $\lambda$4565)& 0.06$\pm$0.02 \\
$F$(N {\sc v} $\lambda$4612)   & 0.14$\pm$0.02 \\
$F$(N {\sc iii} $\lambda$4640) & 0.25$\pm$0.02 \\
$F$(C {\sc iv} $\lambda$4658)  & 0.22$\pm$0.02 \\
$F$(He {\sc ii} $\lambda$4686) & 0.23$\pm$0.02 \\
$F$(C {\sc iii} $\lambda$5696) & 0.10$\pm$0.02 \\
$F$(C {\sc iv} $\lambda$5808)  & 0.15$\pm$0.02 \\ 
\\
$EW$(4650)$^c$           & 4.09$\pm$0.23 \AA \\
$EW$(C {\sc iv} $\lambda$5808) & 1.19$\pm$0.20 \AA \\
\hline
$^a$in units of 10$^{-14}$ ergs s$^{-1}$cm$^{-2}$. \\
$^b$measured in a 6\arcsec$\times$6\arcsec\ aperture \\
\multicolumn{2}{l}{$^c$includes the three WR lines from 4640 -- 4686 \AA.}
\end{tabular}
\end{center}
\end{table}

\subsection{Uncertainties}

Although various empirical methods concur to indicate a similar O/H abundance
for Mkn 309 (cf.\ above), metallicity determinations in the absence of direct 
electron temperature measurements remain uncertain.
This is the case for most of the objects in this study.
The quoted errors indicate only formal uncertainties.

McGaugh (1994) estimates an uncertainty of $\sim$ 0.1 dex on O/H at high abundances.
As shown by Stasi\'nska (1998) calibrations of the strong line method are also
affected by age and geometrical effects. Errors are difficult to 
estimate in this 
case; from theoretical modeling an uncertainty of 0.1--0.2 dex may not be 
unreasonable (Stasi\'nska 1998).

In GIT00 and the present paper we adopt the van Zee et al.\ (1999) calibration
with typical errors of 0.2 dex and which is formally only valid to \oh $<$ 9.1. 
This limited metallicity range is due to the intrinsic difficulties mentioned above
and affects other calibrations in the literature (e.g.\ Edmunds \& Pagel 1984).
The van Zee et al.\ calibration is based on the intensities of [N {\sc ii}] lines.
Our metallicities could be overestimated if nitrogen is enriched
in nuclear regions of galaxies (e.g.\ Coziol \etal\ 1999).

\section{Stellar features and analysis of the stellar populations}

The quality of the spectra allow the detection of several stellar emission
and absorption features the most prominent ones being the WR emission lines, 
H and He absorption lines, TiO, Ca~{\sc ii}, Na~{\sc i} and Fe~{\sc i} 
absorption features.

The broad emission lines attributed to WR stars are discussed below and will
later be used extensively for quantitative modeling. 
The high order Balmer absorption lines as well as \hei\ absorption lines,
often seen in \hii\ regions, are mostly due to OBA stars. 
Recent synthesis models including these lines are given by Gonz\'alez Delgado
\etal\ (1999; cf.\ also Olofsson 1995).

\subsection{Features from late type stars}
Several broad absorption features, mostly due to TiO bands, are observed
around $\sim$ 6250 and $\sim$ 7200 \AA\ in all our objects (see Fig.\ \ref{Fig2}).
These absorption features as well as some narrow absorption lines
of atoms and ions, commonly observed in integrated spectra of clusters
including young stars (cf.\ Bica \& Alloin 1986), indicate the 
presence of red giants or supergiants in our regions. However, some 
contribution of interstellar Ca~{\sc ii} and Na~{\sc i} absorption
might be present.
Quantitative measurements of TiO absorption features for a large sample of
blue compact dwarf galaxies have been provided by Guseva \etal\ 
(1998)\footnote{Using the TiO features, stellar metallicities have been 
obtained for 36 star-forming regions. At least for the lowest metallicities,
 their results
indicate that the metallicities of stars are less than the metallicity of 
the gas.}.
The two absorption features found in our objects are observed in giant and 
supergiant stars of spectral types $\sim$ K3 and later (e.g.\ Silva \& Cornell 1992). 
Despite the presence of nebular lines there is no indication of the broad TiO 
$\gamma^\prime$ band at $\sim$ 6540 -- 7050 \AA\ seen in M types.
The observed TiO features are thus compatible with giants or supergiants of
spectral type $\sim$ K3 or later K types.
Presently we are not able to firmly distinguish between a young ($\ga$ 7--10 Myr)
supergiant population or an older giant population (possibly a ``nuclear'' or bulge
like population).
However, the equivalent widths of heavy element
absorption features are rather consistent with young stellar population.
While strength of TiO bands is insensitive to the age of stellar population
(Bica \& Alloin 1986), the equivalent widths of 
Ca {\sc ii} $\lambda$3933 (1.5 \AA) and Na {\sc i} 
$\lambda$5889, 5895 (1 \AA) lines are smaller than those for old stellar
population and are consistent with values expected for the young stellar
population (Bica \& Alloin 1986).
Therefore, we strongly favour the first interpretation (population of red supergiants),
which, as will be shown later, allows a consistent explanation of all major
observational constraints in the framework of a simple star formation history
dominated by recent activity. Unless otherwise stated this working hypothesis is 
thus adopted in the remainder of this work. 
A more firm distinction should be possible using observations covering a more extended
wavelength range and other stellar signatures.

Limited {\em quantitative} information can be obtained from the observed TiO bands.
First contamination by nebular emission lines and sky lines render exact measurements
of some features difficult. 
Furthermore, quantitative models of the TiO features in young populations are not 
available yet. Although promising for future work (cf.\ Schiavon \etal\ 2000), 
no detailed quantitative analysis of the TiO features will thus be undertaken here.
The only constraint retained below (Sect.\ 5), is that we conclude that 
the RSG features indicate the presence of stars with ages $\ga$ 7--10 Myr in the 
observed region (e.g.\ Origlia \etal\ 1999).

In passing we note the similarity of some of our findings with 
earlier studies on the link between starbursts and AGNs, which have revealed the 
simultaneous presence of the Ca~{\sc ii} triplet originating from red giants or
supergiant, \hii-region like emission lines, and partly also WR stars (e.g.\ 
Terlevich \etal\ 1990, 1996, Pastoriza \etal\ 1993).
The main aim of our study being different from these studies, 
we leave a discussion of similarities and differences to further investigations.

\subsection{Broad emission lines from WR stars -- analysing the WR population}
The objects included in our study have been selected for the presence of
broad emission lines attributed to WR stars. The following lines
have been identified in our objects: \niiib, Si~{\sc iii} $\lambda$4565\footnote{On further
inspection of the data of GIT00 and additional observations, it appears
that the origin of this line (stellar or nebular) may not be fully established.
This will be addressed in a future publication.}, 
the so-called blue bump: \nv, \niii, \civb, \heii;
\ciii, and the red bump \civ\ (see SCK99, GIT00).
The WR line intensities and equivalent widths in Mkn 589, Mkn 710, Mkn 1199 
and Mkn 1259 are taken from GIT00. Additional data on Mkn 710
from SCK99 are also included our analysis. The new data on Mkn 309 is presented below.
The intensities of the blue WR bump in those and other objects 
to be discussed later are also given in Table \ref{Tab5}.
For objects in common with earlier studies, our intensities ($I$(WR)/$I$(\hb)) 
and \whb\ are in good agreement with the data of Mas-Hesse \& Kunth
(1996) and SCK99 for Mkn 710, and with that of Ohyama \etal\ (1997) for Mkn 1259.
Differences with the data of Mkn 710 from Vacca \& Conti (1992) were
discussed in SCK99; their origin remains unclear.

We shall now discuss the observed WR lines and estimate the
number of WR stars of various subtypes in Mkn 309. Such an analysis is given by 
SCK99 and GIT00 for the other objects.

Several broad emission lines (FWHM $\sim$ 15 - 50 \AA)
have been detected in the spectrum of Mkn 309 suggesting the presence of Wolf-Rayet 
(WR) stars of several types.
The observed fluxes of these lines measured in the 2\arcsec$\times$4\farcs2 aperture 
are shown in Table \ref{Tab3}. 
The H$\beta$ flux given there was measured in a  6\arcsec$\times$6\arcsec\
aperture using another spectrum. Comparison with the narrow slit spectrum 
shows that the large aperture contains most of the \hb\ emission which is 
more spatially extended than the continuum emission.

These fluxes have then been corrected for extinction, and line luminosities were 
calculated adopting the distance $D$ = 168.6 Mpc derived from $z$=0.04215 (NED)
assuming $H_0$=75 km s$^{-1}$ Mpc$^{-1}$.
The luminosities have been calculated for two cases: I) the extinction
coefficient $C$(H$\beta$) is equal to 0.68 as derived from the 
Balmer decrement for both gas and stars and II) $C$(H$\beta$) is equal to 0.68
for the gas and 0.15 for stars (Sect.\ 5). 
The results are shown in Table \ref{Tab4}. 
The \hb\ flux and equivalent width agree well with the data of Osterbrock
\& Cohen (1982, hereafter OC82). The intensity of their WR bump is $\sim$ 30 \% 
larger than our measurement including all WR lines from 4512 to 4686 \AA. 
Such differences could have various origins (different observational
setups etc.) and are not uncommon (see e.g.\ comparison in SCK99). They 
do not affect our conclusions on Mkn 309.

The strongest lines are those of the most frequently observed blue 
``WR bump'' ($\sim$ 4650 \AA) which we identify as a composition of 
the N~{\sc v} $\lambda$4603,4619 doublet denoted here as \nv, 
the N{\sc iii} $\lambda$4634-40 blend, the C{\sc iii}/{\sc iv} $\lambda$4650-58 blend, and \heii.
Broad \niiib\ is also detected as previously noted in several objects
by GIT00. 
We confirm the presence 
of \nv, \ciii\ and \civ\ which were suspected by OC82. 

The observed WR features clearly indicate a mixed population of WN and WC
stars. What can be said about their subtypes ? 
The low \nv/\niii\ ratio indicates a late WN subtype; the observed value 
corresponds to the WN6 type, although it is also compatible with subtypes
WN7-8 (Smith \etal\ 1996).
\ciii\ emission is only strong in late WC stars. The observed \ciii/\civ\ 
ratio, the primary classification criterion, corresponds to WC7-8 stars 
(SV98, Crowther \etal\ 1998). Taking into account the contribution
from the detected WNL stars to \civ\ changes the subtype to WC8 according 
to the Crowther \etal\ classification.
The width of \ciii\ (FWHM $\sim$ 19 \AA) is, however, smaller than that 
of \civ\ (49 \AA). This could indicate a mixture of late and some early
WC stars.
Our improved observations confirm the presence of WN and WC stars
which were suspected by OC82 and re-enforce their proposed subtype 
distribution.

%
\begin{table*}[ht]
\caption{Derived parameters for WR lines and stellar populations of Mkn 309}
\label{Tab4}
\begin{tabular}{lcc} \hline
Parameter&I$^a$&II$^b$ \\ \hline
$D$                           & 168.6 Mpc&168.6 Mpc \\
$L^c$(H$\beta$)               & (3.59$\pm$0.28)$\times$10$^{41}$& (3.59$\pm$0.28)$\times$10$^{41}$  \\
$L^c$(N {\sc iii} $\lambda$4512)    & (4.75$\pm$0.49)$\times$10$^{40}$& (1.25$\pm$0.13)$\times$10$^{40}$  \\
$L^c$(Si {\sc iii} $\lambda$4565)   & (1.05$\pm$0.42)$\times$10$^{40}$& (2.82$\pm$1.14)$\times$10$^{39}$  \\
$L^c$(N {\sc v} $\lambda$4612)      & (2.44$\pm$0.43)$\times$10$^{40}$& (6.65$\pm$1.18)$\times$10$^{39}$  \\
$L^c$(N {\sc iii} $\lambda$4640)    &  (4.43$\pm$0.39)$\times$10$^{40}$& (1.21$\pm$0.11)$\times$10$^{40}$  \\
$L^c$(C {\sc iv} $\lambda$4658)     & (3.82$\pm$0.36)$\times$10$^{40}$& (1.05$\pm$0.10)$\times$10$^{40}$  \\
$L^c$(He {\sc ii} $\lambda$4686)    & (4.06$\pm$0.35)$\times$10$^{40}$& (1.13$\pm$0.10)$\times$10$^{40}$  \\
$L^c$(C {\sc iii} $\lambda$5696)    & (1.29$\pm$0.21)$\times$10$^{40}$& (4.71$\pm$0.77)$\times$10$^{39}$  \\
$L^c$(C {\sc iv} $\lambda$5808)     & (1.74$\pm$0.29)$\times$10$^{40}$& (6.52$\pm$1.08)$\times$10$^{39}$  \\
\\
$N$(O7V)                           & 65900$\pm$5100 & 65900$\pm$5100 \\
\\
$N$(WNL)                           & 25800$\pm$2300 & 7060$\pm$630 \\
$N$(WC7) from uncorrected \civ\    & 12400$\pm$2100 & 4650$\pm$770  \\
$N$(WC7) from corrected \civ\ for WNL   
                                   & 10600$\pm$1800 & 4150$\pm$690 \\
$N$(WC7) from \ciii\               & 16000$\pm$2600 & 5810$\pm$950 \\
$N$(WC7)/$N$(WNL)           & 0.42 -- 0.63 & 0.59 -- 0.82 \\ 
\\
$N$(WC8) from \ciii\               & 18800$\pm$3100 & 6840$\pm$1120 \\
$N$(WC4) from \civ\ corrected for WNL+WCL      
                                   &  2780$\pm$460 & 1140$\pm$190 \\
$N$(WC4 + WC8)/$N$(WNL)            & 0.85$\pm$0.14  & 1.13$\pm$0.19 \\
\\
$N$(WNL)/$N$(O7V)                  & 0.39$\pm$0.05  & 0.11$\pm$0.01 \\
$N$(WR)/$N$(O7V)            & 0.59 -- 0.71 & 0.19 -- 0.22 \\
 \hline
\end{tabular}

$^a$$C$(H$\beta$) = 0.68 is assumed for correction of all lines.

$^b$$C$(H$\beta$) = 0.68 and 0.15 are assumed for gaseous and stellar
emission, respectively.

$^c$in ergs s$^{-1}$.
\end{table*}

%
\begin{table*}[htb]
\caption{Properties of related WR galaxies discussed in text (ordered by 
decreasing $L$(WR)).}
\label{Tab5}
\begin{center}
\begin{tabular}{llllllllll} \hline
Name & Metallicity & Distance & $I$(WR)/$I$(\hb)    & $L$(WR) & $N$(WR) & References \\ 
     &  [\oh]      & [Mpc]    & intensity  & [erg s$^{-1}$]  &            \\ \hline
\multicolumn{2}{c}{\em High $L$(WR) objects}\\
IRAS 01003-2238 & & 470 & 0.61& 2.9 $\times 10^{41}$ & 9. $\times 10^4$  & Armus \etal\ (1988) \\
Mkn 309  & 9.3 & 168.6  & 0.34 & 1.2 $\times 10^{41}$ & (1.2--4.) $\times 10^4$ & present work \\
NGC 1614 & 8.6 & 64    & 0.08 & 7.6 $\times 10^{40}$ & 2.4  $\times 10^4$ & Armus \etal\ (1989) \\
 & & & & & & Vacca \& Conti (1992) \\
Mkn 1259 & 8.95 & 28.7 & 0.27 & 5.7 $\times 10^{40}$ & 2.0 $\times 10^4$ & GIT00 \\
Mkn 1199 & 9.13 & 54.1   & 0.38 & 5.9 $\times 10^{40}$ & 1.9 $\times 10^4$  & GIT00 \\
Mkn 477  &     & 152   & $\sim$ 0.1 & 5.4 $\times 10^{40}$ & 1.7 $\times 10^4$ & Heckman \etal\ (1997) \\
F08208+2816 & 8.15 & 189 & 0.08 & 3.7 $\times 10^{40}$ & 2.4 $\times 10^4$ & Huang \etal\ (1998) \\
Mkn 702  & 8.60 & 212.1   & 0.13 & 3.2 $\times 10^{40}$& 1.2 $\times 10^4$  & GIT00\\
Mkn 589  & 8.99 & 45.9    & 0.24 & 2.1 $\times 10^{40}$& 8. $\times 10^3$  & GIT00\\
\\
\multicolumn{2}{c}{\em Objects with large $I$(WR)/$I$(\hb)}\\
Mkn 710 & 9.03 & 20 & $\sim$ 0.3 & 4.6 $\times 10^{39}$ & $\sim$ 800 & SCK99, GIT00 \\ 
Mkn 1063 & 8.46 & 20 & 0.20 & 1.6 $\times 10^{39}$ & $\sim$ 600 & GIT00 \\
Mkn 178 & 7.82 & 2.96 & 0.39 & 1.6 $\times 10^{37}$ & 5 & GIT00 \\
NGC 5128/13& $\sim$ 9. & 5.5 & 1.0--1.8 &  &  &  D'Odorico \& Rosa (1982) \\
\hline
\end{tabular}
\end{center}
\end{table*}

We now estimate quantitatively the number of WR stars using 
the calibration of WR line luminosities of SV98 
from Galactic and LMC stars.
The resulting numbers of WR stars and their relative numbers are shown in Table
\ref{Tab4} for the two choices of extinction coefficient mentioned above.
Since \heii\ is predominantly emitted by WN stars and no nebular emission
is present given the low excitation we use this line to count the number
of WNL stars. Depending on the adopted extinction $\sim$ 7000--26000
stars are found. The number of WC stars can be derived from \civ\ and/or \ciii.
For the former line the contribution of emission from WNL stars 
($\sim$ 10-15 \%) has to be subtracted.
Adopting the mean spectral type to be WC7, we find $\sim$ 4000 -- 16000 WCL.
Strong variations of the line luminosities of individual WC stars on subtype
are found (SV98); adopting WC8 as mean spectral type would considerably
increase the deduced number of WCL stars. 
Assuming that both WCL and WCE stars are present (as indicated by the 
line widths
discussed above), we can alternatively derive their numbers from \civ\ and 
\ciii\ respectively.
The total number of WCL+WCE stars derived in this way is in reasonable
agreement with the previous estimate.
In all cases the resulting $N$(WC)/$N$(WN) ratio is found to be between 
$\sim$ 0.4 to 1.1 (Table \ref{Tab4}).
A metallicity dependence of the average line luminosities used here
can a priori not be excluded. However, Galactic and LMC WR stars do not show 
significant differences, except for \niii\ (cf.\ SV98). 
In addition, since the Galactic objects used by SV98
includes a good fraction of WR stars located inside the solar radius 
at metallicities typically $\sim$ 0.15 dex above solar (cf.\ Maeder \& 
Meynet 1994) our estimated WR populations and the detailed comparisons 
shown below should not be affected by possible variations of the 
properties of individual WR stars with metallicity.

What can be said about the remaining WR lines ? Using
 again the line luminosities
from SV98, different estimates of WC stars yield predictions for the \civb\ 
line in good agreement with the observed one (differences from --20 to +50 \%).
The observed \niii\ emission from WNL stars is found to exceed the expected value
by a factor of $\sim$ 2 based on the average Galactic value of \niii/\heii\ 
(see SV98), which is known to show a metallicity dependence (Smith \etal\ 1996).
The excess could thus be due to larger \niii/\heii\ emission in WNL stars
at high metallicities (cf.\ also Schmutz \& Vacca 1999).
Using the derived WR number population and the data of SV98, weak emission --
below our detection limit -- is expected for other WR lines. This is in agreement
with the observed spectrum.
Similar quantitative comparisons cannot yet be done for \niiib\ and
Si~{\sc iii} $\lambda$4565 (see GIT00 for detections of these 
lines in other objects).
We conclude that all broad emission lines detections and non-detections are 
compatible with the WNL and WC populations indicated in Table \ref{Tab4}.

\begin{figure}[htb]
\centerline{\psfig{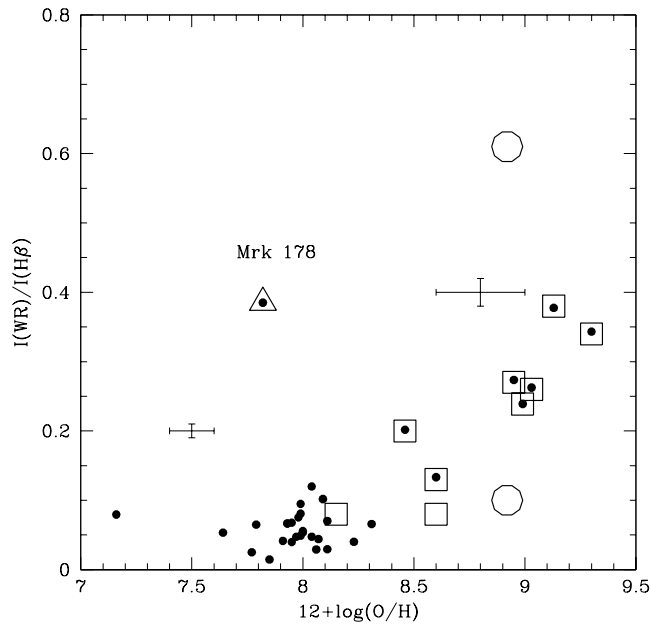}}
\caption{Observed $I$(WR)/$I$(\hb) ratio for WR galaxies from GIT00 (filled dots) and 
WR galaxies from Table \protect\ref{Tab5} (open squares and circles) as 
a function of metallicity. The low $L$(WR) object Mkn 178 is marked separately 
(open triangle).
Typical uncertainties are shown by the errorbars. The formal uncertainties
are $\le$ 5 percent for the line intensities from GIT00. 
For objects with \oh\ $\le$ 8.3 (measured from [O~{\sc iii}] $\lambda$4363)
the uncertainties on the metallicity are $<$ 0.1 dex. Larger uncertainties are 
expected for the remaining objects (cf.\ Sect.\ \protect\ref{s_abund}).
NGC 5128/13 with a very uncertain $I$(WR)/$I$(\hb) measurement is not shown.
Circles indicate objects of unknown metallicity, placed arbitrarily at solar
metallicity.
The $I$(WR)/$I$(\hb) ratios are {\em uncorrected} for aperture effects (cf.\ GIT00) and 
possible differential extinction between gas and stars.
The large $I$(WR)/$I$(\hb) ratio of Mkn 178 is likely due to 
an exceptionally large aperture effect for which GIT00 estimate a reduction
of a factor 5 or due to small number statistics.}
\label{Fig3}
\end{figure}

\subsection{WR and O star populations in Mkn 309 and related objects}

The ionizing flux of Mkn 309 corresponds to a number $N$(O7V)$\approx$65900
of equivalent O7V stars 
assuming Case B recombination, $T_e=5000$ K, and 
$n_e=100$ cm$^{-3}$.
Thus, the number ratio of WR stars to equivalent O7V stars
 is $N$(WR)/$N$(O7V) = 0.59 -- 0.71 in case I (Table \ref{Tab4}) and 0.19 -- 0.22 
in case II.
Here $N$(WR) = $N$(WNL+WC4+WC8) or $N$(WR) = $N$(WNL+WC4+WC7).
The high $N$(WR) / $N$(O7V) value (assuming identical extinction for gas and stars) 
is more than a factor of 2 larger than the observed $N$(WR)/$N$(O) value in Galactic 
regions inside the solar radius and in M31, which are representative for 
environments with constant star formation at solar and supersolar metallicity 
(see Maeder \& Meynet 1994 and references therein).
As for other WR galaxies (cf.\ e.g.\ Vacca \& Conti 1992, GIT00) such 
$N$(WR)/$N$(O) 
ratios generally indicate intense star forming events taking place over 
relatively short timescales (e.g.\ review of Schaerer 1999a). 
Estimates of the SF timescales for the objects of our study and allowing
also for the observed lower extinction for the stars will be derived 
below (Sect.\ 5).

Among all WR galaxies with \oh\ $\ga$ 8.3, the largest $N$(WC)/$N$(WN) ratio
($\sim$ 0.4 -- 0.8) is derived in Mkn 309 (cf.\ SCK99, GIT00). 
Although lower, the upper value comes close to $N$(WC)/$N$(WN) $\sim$ 0.9 -- 1
observed at high metallicities in the Local Group (Massey \& Johnson 1998).
As shown below, the WN and WC populations in all metal-rich WR galaxies studied
here can be understood in terms of current stellar evolution models.

To compare the WR populations of Mkn 309 with those in other related objects
we have compiled in Table \ref{Tab5} the properties of WR galaxies with
large WR bump luminosities and large $I$(WR)/\hb\ line intensities.
The compilation is based on the catalogue of Schaerer \etal\ (1999b) and the data
discussed in Schaerer (1999a).
In absolute terms, the WR population in Mkn 309 is among the largest
known; presently it is only surpassed by the ultra-luminous infrared galaxy
IRAS 01003--2238 ($\equiv$IRAS F01004--2237, Armus \etal\ 1988, Veilleux \etal\ 1999). 
Other objects with more than $\sim$ 8000-10000 WR 
stars\footnote{WR numbers based on determinations in the original papers 
or objects with $L$(WRbump) $\ga$ 2.$\times 10^{40}$ erg s$^{-1}$
using the calibration of Smith (1991) for individual WN7 stars.
The list is likely incomplete due to missing data on WR line fluxes.
Marginal cases not included here are NGC 6764, Mkn 315, He 2-10
(references in Schaerer \etal\ 1999b). 
See GIT00 (Fig.\ 3) for additional WR-bump luminosities.} 
are NGC 1614, 
Mkn 477, 
Mkn 1259, Mkn 1199,
Mkn 702 and Mkn 589, 
and F08208+2816 
whose main properties are summarised in Table \ref{Tab5}.

Interestingly, at least the two most ``WR-luminous'' objects (IRAS 01003--2238 
and Mkn 309) also stand out by very large $I$(WR)/$I$(\hb) 
ratios\footnote{$I$(WR) 
stands for the intensity of blue bump including broad lines between 4640 and 
4686 \AA.} as shown in Fig.\ \ref{Fig3}. 
The intensities (see Table \ref{Tab5})
clearly exceed the typical values of $\la$ 0.1--0.2 (cf.\ also Schaerer 1999a).
Other objects with strong WR bump intensities but much {\em smaller
absolute populations} of WR stars are the \hii\ region \# 13 in NGC 5128 
($\equiv$Cen A), Mkn 178, Mkn 710,
and Mkn 1063 also 
listed in Table \ref{Tab5}.
Although few such WR regions 
are currently known at the high 
metallicity end, it seems
that the largest WR intensities are found in 
massive and distant objects. If true, this is quite surprising.
Indeed for similar metallicity and IMF one would statistically expect 
{\em reduced WR strengths} in more distant objects where the spectroscopic 
observations encompass a much larger area containing more likely a mixture of 
several populations with different ages. 
Observational biases may, however, also be present (see 
Kunth \& Joubert 1985).
New observations of nearby (low $L$) metal-rich WR regions will be 
needed to confirm this trend. 
In any case, it appears that despite the likely growing importance of the 
``age spread effect'' the increase of the WR population with metallicity 
(cf.\ Maeder \& Meynet 1994) is sufficient to lead to the continuing
increase of the $I$(WR)/$I$(\hb) envelope with metallicity first shown by 
Arnault \etal\ (1989).

\section{Model comparison} 

To constrain the basic properties of the massive star populations
(IMF slope, \mup) and the star formation history (age, burst duration)
of our objects we now undertake a detailed comparison between the
observations and evolutionary synthesis models.

\subsection{Procedure}

The following main observational constraints are used:
\begin{enumerate}
\item {\em \hb\ and \ha\ equivalent widths.} The former is used as a primary 
  age indicator; once \whb\ is reproduced \wha\ serves as an independent
  test for the predicted spectral energy distribution (SED) in the red.
\item {\em Nebular line intensities.} $F$(\ha)/$F$(\hb) determines the extinction
  of the gas. The use of other line intensities requires detailed photoionization
  modeling which is beyond the scope of this paper.
\item {\em Intensities and equivalent widths of the main WR features.}
  The blue bump (henceforth referred to as 4650 bump) and \civ\ (red bump)
  serve as main constraints on the WR population. To avoid uncertainties
  in deblending individual contributions of the blue bump we prefer to
  use measurements for the entire bump. 
  In contrast to the spectra of metallicity objects our low excitation spectra show 
  no evident contamination from nebular lines (e.g.\ [Fe~{\sc iii}] $\lambda$ 4658,
  nebular \heii).

  To potentially disentangle between various effects (underlying ``non-ionizing'' 
  population, loss of photons, differential extinction between gas and stars)
  it is important to use both equivalent widths and relative WR/\hb\ intensities
  (cf.\ SCK99).
\item {\em TiO bands at $\sim$ 6250 and 7200 \AA}. 
  Assuming that they originate from a red supergiant population (rather than giants;
  cf.\ Sect.\ 4) this indicates the presence of cool stars
  from a population with ages $\ga$ 7--10 Myr.
\item {\em SEDs} available over the full range of $\sim$ 3500--7500 \AA\ provide
  an additional important constraint on the non-WR populations, which are responsible
  for the continuum flux. 
\end{enumerate}

For the model comparisons we use calculations based on the evolutionary synthesis code 
of SV98, which in particular includes the most recent calibration of WR line
luminosities used to synthesize the WR features, up-to-date stellar tracks, and recent 
stellar atmospheres for O and WR stars complemented by Kurucz models for cooler stars
(see SV98 for a full description). In all cases the high-mass loss stellar tracks
of Meynet \etal\ (1994), which reproduce a large number of properties of individual
WR stars and WR populations in nearby galaxies (Maeder \& Meynet 1994) are used.
Note that except for the improved O star atmospheres used by SV98 the recent 
{\em Starburst99} synthesis models (Leitherer \etal\ 1999) use the same basic 
input physics.

\begin{figure*}[htb]
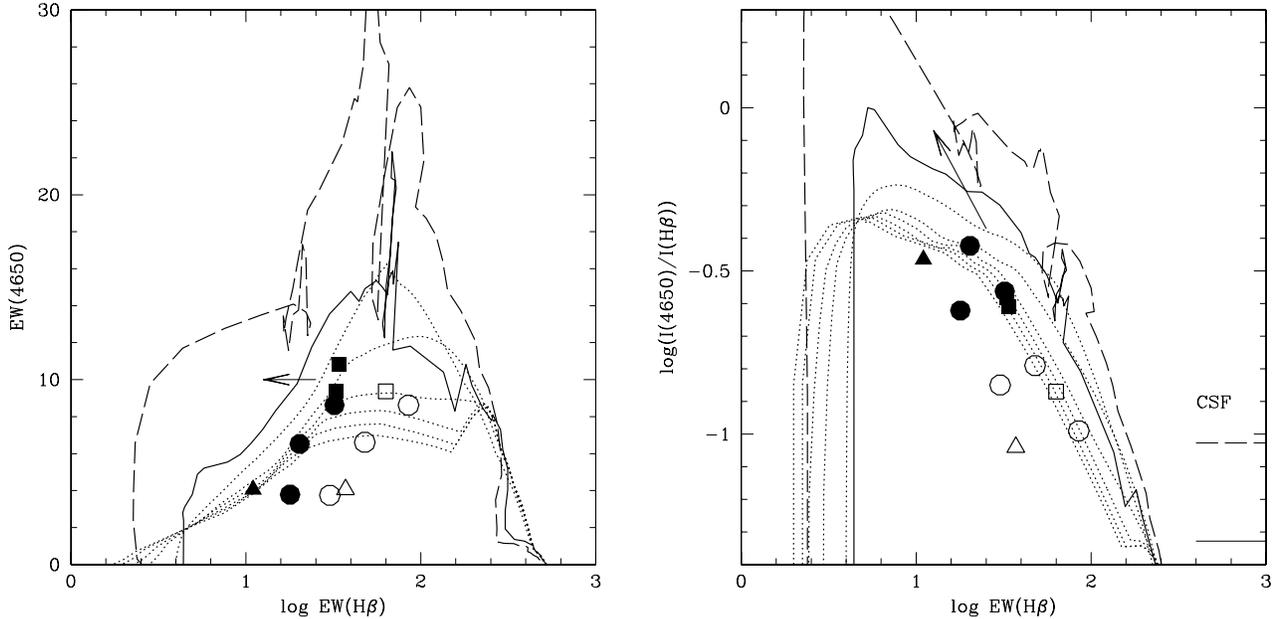

\centerline{\psfig{file=9705.f4a,width=8.8cm}
            \psfig{file=9705.f4b,width=8.8cm}}
\caption{\label{Fig4}
Observed and predicted equivalent width (left panel) and line intensity
with respect to \hb\ (right panel) as a function of \whb. Observed points are:
Mkn 309 (triangle), Mkn 589, Mkn 1199, Mkn 1259 
(circles), two measurements for Mkn 710 ($\equiv$NGC 3049) from SCK99 and GIT00 (squares).
Filled symbols assume a unique extinction correction for gas and stars;
open symbols include differential stellar-gaseous extinction (see text).
Data are uncorrected for aperture effects (cf.\ GIT00).
Typical uncertainties are: 5--10 \% for \whb, $\le$ 10 \% for $EW($WRbump$)$,
and $\sim$ 0.05 dex in $\log(I($WR)/$I($\hb)).
Model predictions are shown for instantaneous bursts at $Z$=0.02 (solid line)
and $Z$=0.04 (dashed line), and extended bursts at $Z$=0.02 (dotted
line; burst durations
$\dt=2,4,6,8,10,12$ Myr).
All models assume a Salpeter IMF with \mup=120 \msun.
The arrow illustrates the shift which has to be applied to the model predictions
in case of a decrease of \hb\, e.g.\ due to absorption of photons by dust
(see text).
The observed WR bump is well reproduced by extended bursts with durations
$\dt \sim$ 4--10 Myr.
}
\end{figure*}

The basic model parameters we consider are:
\begin{enumerate}
\item[a)] {\em Metallicity.} Two sets of stellar tracks with different metallicities are 
  explored here: $Z=0.02$ (solar) and 0.04.
\item[b)] {\em IMF slope and upper mass cut-off (\mup).} We adopt a Salpeter IMF
  (slope $\alpha$=2.35), and \mup=120 \msun\ as our standard model.  
\item[c)] {\em Star formation history (SFH).} Models for instantaneous bursts
  (coeval population), extended burst durations (constant SF during period \dt; 
  in this case age=0 is defined at the onset of SF, i.e.\ corresponds to that of
  the oldest stars present),
  and constant SF are considered. Models of combined stellar populations (arbitrary ages,
  and relative weights) with the same SFH have also been calculated.
\item[d)] {\em Fraction of ionizing Lyman continuum photons ($f_\gamma$).} 
   $f_\gamma$ indicates the fraction of ionizing photons absorbed by the gas.
   Our standard value is $f_\gamma=1$.
   Values $f_\gamma < 1$ are used to simulate various effects (e.g.\ dust absorption, photon
   leakage outside regions, etc.) leading to a reduction of photons available for
   photoionization.
\item[e)] {\em Stellar extinction} is kept as a free parameter for the comparison
  of the SED. This is justified by numerous indications showing different extinction
  between gas and stars (e.g.\ Fanelli \etal\ 1988).
  Note that such a difference affects also \wha, \whb, and $I$(WR)/$I$(\hb) ratios.
  In Figs.\ 4, 5, 8 we therefore show two sets of observations: corrected assuming
  $C(\hb)_\star=C(\hb)_{\rm gas}$ (filled symbols) and using $C(\hb)_\star$ (open symbols)
  determined from the SED model fit (see below).
\end{enumerate}

\begin{figure}[htb]
\centerline{\psfig{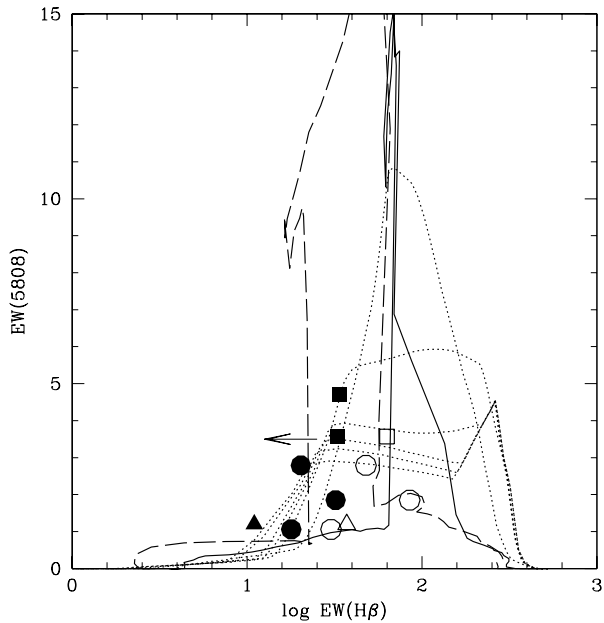}}
\caption{\label{Fig5}
Same as Fig.\ \protect\ref{Fig4} for the equivalent width of the 
\civ\ WR bump.}
\end{figure}

One of our main aims is to constrain the IMF (slope, \mup). To do so
we adopt in this section a ``conservative approach'' which consists of exploring 
as far as possible a ``standard'' (Salpeter) IMF to see if any observational 
constraint requires a deviation from it.
Regarding the WR features this in particular implies 
that in this section (i.e.\ Figs.\ \ref{Fig4} and \ref{Fig5})
all observations will be compared to synthesis models at solar metallicity ($Z$=0.02) 
This model can thus
be taken as a {\em lower limit} for the predicted WR strengths which increase with
increasing metallicity. The use of this lower limit is necessary in view of the
lack of precise metallicity determinations for metal-rich objects.

\begin{figure}[tb]
\centerline{\psfig{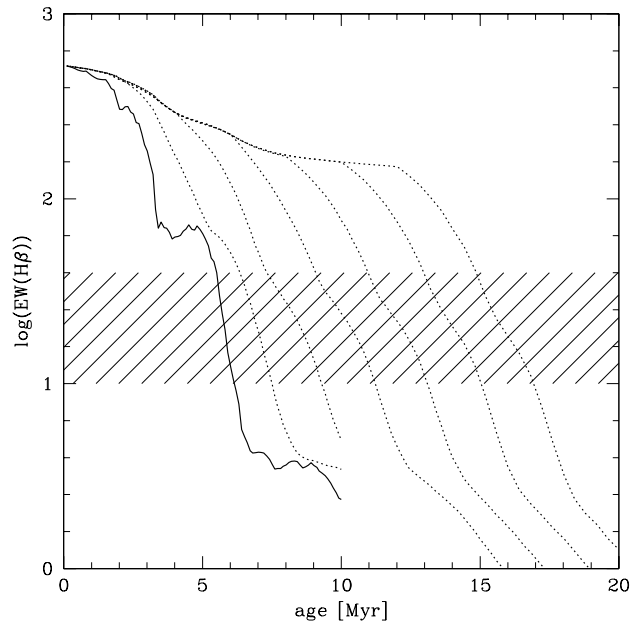}}
\caption{\label{Fig6}
Temporal evolution of the predicted \hb\ equivalent width for the solar metallicity
(Z=0.02) models shown in Figs.\ \protect\ref{Fig4} and \protect\ref{Fig5}.
Solid line: instantaneous burst, dotted lines: extended bursts with durations 
$\dt=2,4,6,8,10,12$ Myr.
For Z=0.04 the same equivalent widths are obtained at ages on average 
$\sim$ 1 Myr younger.
The shaded band shows the range of the observed equivalent widths.
}
\end{figure}

\subsection{Results}
The comparison of the observed and predicted WR features is shown in Fig.\ 
\ref{Fig4}. We show predictions for instantaneous bursts at $Z$=0.02 (solid
line), $Z$=0.04 (dashed), and burst durations of $\dt = 2,4,6,8,10,12$ Myr at $Z$=0.02
(dotted). All models are calculated with the ``standard'' IMF (Salpeter, 
\mup=120 \msun).
Figure \ref{Fig4} shows the following:

\begin{itemize}
\item both the observed WR equivalent width and the 
$I$(WR) / $I$(\hb) ratio fall below
the predictions for an instantaneous burst at solar metallicity rather than 
between the $Z$=0.02 and 0.04 curves corresponding to super solar metallicities.
If each observed region corresponded indeed to a single coeval population, 
this finding would be a strong argument in favor of a deficiency of stars
above the WR mass limit ($M_{\rm WR} \sim$ 21--25 \msun) with respect to the
standard IMF.
However, in the present case this can be explained quite simply as follows.

\item The observations are well reproduced by $Z$=0.02 models assuming extended
burst durations of $\dt \sim$ 4--10 Myr.
The corresponding ages of our objects, as indicated by \whb, are between 
$\sim$ 7 and 15 Myr as shown in Fig.\ \ref{Fig6}.
The observations of the red WR bump (\civ; see Fig. \ref{Fig5}) and the remaining
observational constraints are well matched by this scenario (see below).

\item Including the differential extinction correction for gas and stars as derived
from the Balmer decrement and the SED (cf.\ below) leads to non-negligible changes
of the considered quantities. Allowing for this effect does, however, not modify
the above conclusions for the bulk of the objects.

\end{itemize}

The extended burst scenario at ages $\sim$ 7--15 Myr also explains the presence 
of red (super)giants revealed by the TiO bands (Sect.\ 4).
Even at a quantitative level we note, somewhat surprisingly, that the broad TiO features 
(especially the 6250 \AA\ feature) in the majority of objects are quite well 
reproduced by our SEDs, where Kurucz models are used to describe cool stars
(see Fig.\ \ref{Fig7}).
In any case, we find that the overall shape of the SED alone already indicates 
a mixture of populations with different ages. Independent of the amount of 
extinction, single stellar populations are not able to produce the required 
relative amount of blue ($\la$ 5100 \AA) to red light.

Last, but not least, the overall SEDs in the optical range of all objects
are very well reproduced with the extended burst models at $Z$=0.02
\footnote{Similar agreement is also obtained with the $Z$=0.04 models.}.
The SEDs, shown in Fig.\ \ref{Fig7}, correspond to the model for \dt=8 Myr,
taken at the age deduced from the observed \hb\ equivalent width (see 
Fig.\ \ref{Fig6}).
Both the predicted stellar and nebular continuum emission are included in 
the models. In the present case the latter is of importance only for emission
shortward of the Balmer break. However, in this domain a quantitative comparison 
is not possible since the H and He continuous emission coefficients depend
quite strongly on the electron temperature which is poorly known
for the present objects.

As discussed above, the extinction suffered by the stellar continuum ($C(\hb)_\star$)
is taken as a free parameter, since it may differ from the value derived from 
the gas (Balmer decrement). The resulting values of $C(\hb)_\star$ are systematically 
{\em smaller} than the values from the gas
(Mkn 309: $C(\hb)_\star=$ 0.15 / $C(\hb)_{\rm gas}=$0.68, Mkn 589: 0.5/0.72; 
Mkn 710: 0.25/0.52, Mkn 1199: 0.2/0.55, Mkn 1259: 0.2/0.60) with differences
similar to those found e.g.\ by Calzetti (1997) and Mas-Hesse \& Kunth (1999).
For Mkn 309, with the largest $\Delta C(\hb)$, we have independent evidence 
showing that $C(\hb)_\star < C(\hb)_{\rm gas}$:
the observed ratio \wha / \whb = 8.9~ clearly exceeds the maximum (typical) value
of $\sim$ 7.5 ($\sim$ 5--6) predicted for instantaneous (extended) burst 
models\footnote{Although less clearly the remaining objects seem to indicate 
the same finding. 
\wha/\whb\ is larger than the typical value predicted for extended burst models.
Mkn 589: \wha/\whb=6.30; Mkn 710: 6.68; Mkn1199: 6.84; Mkn 1259: 7.18.}.
Any other effect (underlying old population, multiple components etc.) will 
further reduce the predicted \wha/\whb\ ratio; the only way to bring both in 
agreement is by adopting a larger extinction for the gas.
Accounting for the differential extinction and adopting the SEDs shown in Fig.\
\ref{Fig7}, the observed \wha / \whb\ are in excellent agreement with the model
predictions for all objects.

\begin{figure*}
\centerline{\psfig{file=9705.f7,width=15.0cm}}
\caption{\label{Fig7}
Comparison of observed spectral energy distributions with predicted ones
for extended burst models (\dt=8 Myr) with a standard IMF.
All SEDs are normalised at 4800 \AA. The predicted SEDs, which do not include
emission lines, are reddened by the variable amount $C(\hb)_\star$ indicated 
in each panel. 
}
\end{figure*}

We conclude that all the given observational constraints can be reproduced
by models with a Salpeter IMF extending to high masses for a burst scenario with
star formation extending over $\sim$ 4--10 Myr.
Let us now address the question if this model is unique and what other models 
can be excluded.

\subsection{Alternate models}
\label{s_alt}
A variety of other models have been considered to explain the observations. 
We here consider mostly three variations, all of them adopting a large
upper mass cut-off:
{\em 1)} the effect of internal dust absorbing the ionizing photons,
{\em 2)} variations of the star formation history, and
{\em 3)} variations of the IMF slope.

The first one is illustrated in Fig.\ \ref{Fig4}, where the arrow indicates
the direction along which the {\em predicted} lines will be shifted if 50 \% of the 
Lyman continuum photons are lost (e.g.\ absorbed by dust inside the \hii\ region).
The Figure shows that in this case the observations can still reasonably be 
explained with our ``standard model''; somewhat longer burst durations are
then required. 
A distinction between the two cases from the remaining constraints including the
SED is not possible. However, values of $f_{\gamma} \la 0.5$ seem 
more difficult to reconcile with Fig.\ \ref{Fig4}.

As mentioned earlier instantaneous burst models are excluded from the 
observations due to the presence of both WR and RSG features, and the detailed SED.
This also holds for synthesis models allowing for the formation of WR stars
through Roche lobe overflow in massive close binary systems
(Cervi\~no 1998, Cervi\~no \etal\ 2000).
At the epoch where such instantaneous burst models predict the simultaneous
presence of WR (formed in interacting systems) and RSG, the predicted
\hb\ equivalent width is too low (Cervi\~no, private communication).
Constant star formation (CSF) predicts a maximum $I$(WR)/$I$(\hb) which is much 
too small (see Fig.\ \ref{Fig4}), and too large \hb\ equivalent widths.
The only way reconcile this scenario is to invoke effect 2) with variable values 
of $f_\gamma$ as low as $\sim$ 0.1--0.25. The SED does not exclude this possibility.
We have also performed Monte-Carlo simulations combining arbitrarily two ``standard IMF'' 
instantaneous bursts of different ages and relative contributions. Several possible 
combinations are found satisfying all the observational constraints; all of them 
obviously involve a young ($\la$ 5 Myr) population superposed to a older one 
($\ga$ 7--9 Myr) producing enough evolved red stars. Again, no strong distinction 
is found (``by eye'') for the SEDs compared to the extended burst scenario discussed 
above.

Finally we have considered the case of a steeper IMF slope of $\alpha=3.3$,
keeping \mup=120 \msun. This IMF deficient in massive stars with respect to
the Salpeter slope produces just a strong enough WR bump (up to $EW($4650) $\sim$ 
9 (13) \AA\ for $Z$=0.02 (0.04)) at the maximum during an instantaneous burst.
Extended bursts are essentially excluded in this case, and very few solutions 
(combined populations) can be found. Not enough massive stars (WR stars with
$M_{\rm ini} \ga$ 20--25 \msun) are expected for this steep IMF slope.

From the models considered in this Section we conclude that a variety of models 
with a ``standard IMF'' (Salpeter slope and large \mup) reproduce the observational 
constraints.
Very few solutions can be found for an IMF slope $\alpha =3.3$ close to the Miller \& Scalo
(1979) value; this renders such a steep IMF in metal-rich regions very unlikely.

Before we proceed to a further discussion of the IMF we shall briefly compare
our successful models with those available by other authors for Mkn 710 
and Mkn 1259.

\subsection{Comparison with other work}
The WR region in Mkn 710 
has recently been studied in detail by 
SCK99 and Mas-Hesse \& Kunth (1999).
Our results are in agreement with the study of SCK99, whose measurements have
been included in the present paper (see also SCK99 for comparisons with the 
earlier data of Vacca \& Conti 1992).
This region being the only metal-rich 
in the sample of SCK99, no detailed fitting attempt was made. 
Regarding the IMF, SCK99 conclude that the Salpeter slope produces enough
WR stars to explain the observations. From the analysis of 5 metal-poor 
regions SCK99 found that short burst durations ($\dt <$ 2--4 Myr) were required;
Mkn 710 was not considered. The somewhat longer burst durations found for the 
present metal-rich sample could be due to an increasing complexity of the regions
considered here (nuclear SB versus isolated \hii\ regions) hosting progressively
more separate populations.

For the comparison with Mas-Hesse \& Kunth (1999) it is important to note
that their optical--UV observations include a much larger region (10 by 
20\arcsec)
than the present observations typically extracted over 2--3\arcsec\ wide regions.
This explains most likely the larger contribution of an older underlying population
contributing to the optical flux found by these authors. The results on the burst
duration are in good agreement. Incidentally Mas-Hesse \& Kunth derive the same 
stellar extinction as found from our work. 

For Mkn 1259 a subsolar metallicity was derived by Ohyama \etal\ (1997) based on
[O {\sc iii}] $\lambda$5007\AA. The value used in the present work, based on a spectrum 
of superior S/N and resolution by GIT00, should be more accurate.
Using their low metallicity value, Ohyama \etal\ (1997) concluded that an instantaneous burst 
with an age of $\sim$ 5 Myr reproduces well the observed properties 
of the nuclear region of this object.
Using our metallicity, the strengths of the WR features imply
an extended burst, compatible with the observed RSG features and the SED.
For most aspects however, the difference with Ohyama \etal\ (1997) is probably negligible.

\section{Discussion} 

Having excluded steep IMF's with a high \mup, we will now examine what limits 
can be set up on \mup\ for a Salpeter slope and discuss several implications
from our results.
We then briefly discuss the plausibility of the star formation timescales derived 
in this work.

\begin{figure}[tb]
\centerline{\psfig{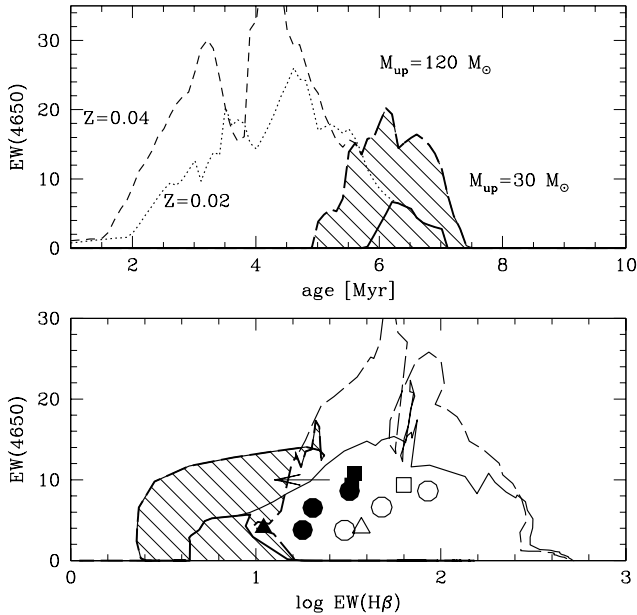}}
\caption{\label{Fig8}
Predicted dependence of the equivalent width of the WR bump on variations
of the upper mass cut-off from \mup $=$ 120 \msun\ (thin lines) to \mup\ $=$ 30
\msun\ (thick lines).
Instantaneous burst models with a Salpeter IMF for solar metallicity (dotted,
solid) and $Z$=0.04 (short-dashed, long-dashed) are shown.
The shaded area indicates the maximum domain covered by twice solar metallicity models
with the reduced mass cut-off (\mup\ $=$ 30 \msun).
{\em Upper panel:} Temporal evolution of $EW$(WR). 
{\em Lower panel:} Predicted $EW$(WR) as a function of $EW$(\hb). The observations
are shown using the same symbols as in Fig.\ \protect\ref{Fig4}.
The strong dependence on \mup\ illustrates the power of using the WR features
to constrain \mup. Discussion in text.}
\end{figure}

\subsection{Constraints on the IMF from the present data}
{\em What value of \mup\ is our data exactly sensitive to ?}
From the ages and burst durations derived above we see that the {\em youngest
stars} present have ages of the order of $\sim$ 3--5 Myr. 
This corresponds to stars with initial masses $M_{\rm ini} \sim$ 40--120 \msun\
(see Meynet et al.\ 1994). 
Although in the best case our data could be sensitive to changes of \mup\ over this 
range, we have to conservatively admit that we might not be able to distinguish
between different values for the upper mass cut-off $\ga$ 40 \msun.
To improve the present situation one obviously has to be able to study the
intrinsically youngest regions and/or include additional observational constraints.

With the objects studied here it is not straightforward to answer
the reverse question, namely {\em ``What lower limit can we set on \mup ?''}.
In the ideal case (instantaneous burst with a known metallicity) the use of 
WR lines is a very powerful discriminant, as illustrated in Fig.\ 8.
Here, in principle, the high metallicity tracks ($Z$=0.04) should be adopted
to obtain a conservative lower limit for \mup.
However, as it is clear from Sect.\ 5, the use of the WR features is
not straightforward for answering the above question given the extended duration
of SF. 
On the other hand, as apparent from Fig.\ \ref{Fig8}, the observed \hb\ equivalent 
widths indicate clearly an upper mass cut-off of \mup\ $>$ 30 \msun.

The $I$(He~{\sc i} $\lambda$5876)/$I$(\hb) ratio has previously been used by Bresolin
\etal\ (1999) as an indicator of \mup.
We note that for all of the data except Mkn 309 our limit on \mup\ seems also 
consistent with the observed $I$(He~{\sc i} $\lambda$5876)/$I$(\hb) ratio as estimated 
from simple Case B recombination theory. 
More firm conclusions await combined stellar population and photoionization modeling 
(cf.\ Sect.\ 7).

\subsection{IMF constraints from \ha/\hb\ equivalent widths of young SF regions}
Assuming a Salpeter IMF with \mup\ as large as $\sim$ 100--120 \msun\ at high 
metallicity, to be compatible with the observations discussed above, 
has obviously implications which need to be confronted with other observational data.
A straightforward implication is that at age=0 \hii\ regions are expected
to show very large \hb\ equivalent width ($\sim$ 450-- 500 \AA\ for 
$Z$=0.02--0.04, and larger values for $Z$ $<$ $Z_\odot$ and \mup=120 \msun).
It is well known that regions with such large \whb\ are apparently not 
found or are extremely rare, independently of metallicity to first order. 
What is the situation for high metallicities ?
Relatively few \whb\ measurements have been published for metal-rich \hii\ regions.
Among the data of Shields \etal\ (1991), Oey \& Kennicutt (1993), Kennicutt \&
Garnett (1996), van Zee \etal\ (1998), and Bresolin \etal\ (1999, see Bresolin
1997) few regions with 
\oh\ $\ga$ 8.9 have \whb\ $\ga$ 250 \AA. Furthermore these measurements
are most likely affected by very large uncertainties due to placement of
the weak continuum as inspection of some data of van Zee \etal\ shows.
From this data we conclude that max(\whb) $\sim$ 200-300 \AA\ for metal-rich 
\hii\ regions.
For a Salpeter slope this corresponds to \mup\ $\sim$  35 -- 50 \msun,
compatible with our conservative lower limit for \mup\ but clearly 
below ``normal'' values of $\sim$ 100--120 \msun.

Although a detailed discussion and interpretation of the observed \ha\ or 
\hb\ equivalent widths distributions is beyond the scope of the present
work, we would like to briefly comment on four processes which could 
lead to reduced values of the observed \whb\ maximum:
{\em 1)} Dust inside the \hii\ regions, 
{\em 2)} underlying populations diluting the continuum,
{\em 3)} early phases hidden in compact radio sources (cf.\ Kobulnicky \& Johnson 
  1999), 
{\em 4)} absence of a proper ZAMS for high mass stars in the accretion scenario
  of Bernasconi \& Maeder (1996).
From the available data we are not able to discuss possible selection effects.

The first two are the most commonly considered possibilities. The importance of dust,
especially in metal-rich objects, remains to be explored. For metal-poor objects
(e.g.\ \hii\ galaxies) with a low dust content the effect should on average be
small. 
Although for metal-poor objects the young bursts might dominate the visible light 
and underlying absorption (2) should thus not be important for the max(\whb)-discrepancy 
(cf.\ Mas-Hesse \& Kunth 1999), the recent study of Raimann \etal\ (2000) suggests
the existence of non-negligible underlying populations.
If the bulk of star formation (SF) occurred in a similar fashion as in He 2-10,
where Kobulnicky \& Johnson (1999) find a significant number of compact radio
sources, this could constitute a possibility to hide the earliest phases of SF
and thereby reducing the observed maximum \whb.
We note, however, that the estimated lifetime of these regions ($\sim$ 10 \%
of the O star lifetime; Kobulnicky \& Johnson) is clearly too short to resolve
the max(\whb)-discrepancy which concerns the first $\sim$ 2 Myr.
Effect 4 has been suggested by Mas-Hesse \& Kunth (1999) as a possibility
to explain this discrepancy.
However, test calculations (Schaerer \& Bernasconi, 1995, unpublished) show that
the stellar tracks following the accretion scenario of Bernasconi \& Maeder (1996)
lead to an insufficient decrease of max(\whb)\footnote{max(\whb) $\sim$ 400 \AA\ for
$Z$=0.02 and the standard IMF.}.
In addition, the validity of their scenario remains to be verified by detailed 
confrontations with observations of individual stars.

Regarding only metal-rich region we conclude from the available \whb\ data 
that the observed maximum \hb\ equivalent width is compatible with a 
Salpeter IMF with an upper mass cut-off \mup\ $\sim$ 35 -- 50 \msun,
in agreement with the results from our WR galaxies.
Larger values of \mup\ cannot be excluded from our present (small) sample 
of objects, in contrast to the indications from \whb\ measurements
in larger samples.
In view of the discrepancy between the observed and predicted max(\whb)
which seems to hold {\em for all metallicities} and for which currently 
no clear explanation exists, interpretations based on the ``exclusive'' 
use of \whb\ should, however, be taken with caution.

\subsection{Are the derived burst durations plausible ?}
An interesting result derived from our models is that all the objects 
analysed here require extended burst durations over $\sim$ 4--10 Myr
or a superposition of two or more bursts with similar ages differences and
a young component ($\la$ 5 Myr).
\footnote{
We note that the observed \hb\ equivalent width and the
$I$(WR)/$I$(\hb) ratio of IRAS 01003--2238 (cf.\ refs.\ in Table \ref{Tab5})
can be explained by $Z$=0.04 models with a longer burst 
($\Delta t \sim$ 20 Myr).}
For the most distant object from our sample (Mkn 309) the size of the 
observed region corresponds to $\sim$ 2 kpc. For a typical galaxian 
velocity dispersion of 200 km s$^{-1}$ the crossing time is 8.4 Myr.
Shorter times are obtained for the remaining objects (cf.\ Heckman \etal\ 
1997 for Mkn 477) including also IRAS 01003--2238, which has a 
very compact morphology (Surace \etal\ 1999).
This simple estimate shows that the above burst durations 
pose no fundamental physical problem for any of the objects discussed here.

Is there any difference of increasing burst duration with distance as could 
be expected on average if the observed regions are composed of an increasing number
of ``small'' super star cluster like objects ?
In this context it is interesting to note that in an earlier study of WR galaxies 
with distances of 4--14 Mpc
burst durations not exceeding $\sim$ 2--4 Myr were found using the same evolutionary 
synthesis models (SCK99). This difference is likely due to two effects:
first the distance effect, and second the different nature of the objects
included in the two samples (dwarf galaxies with one or few SF regions versus
predominantly nuclear starbursts).

\section{Summary and conclusions} 

We have obtained new spectroscopic observation of Mkn 309, a starburst galaxy 
with one of the largest WR population known (cf.\ Osterbrock \& Cohen 1982, 
Armus \etal\ 1988). 
Our high S/N observations allow in particular a detailed determination 
of the WR star content of this object, composed of WNL, WCL and possibly 
also some WCE stars (Sect.\ 4).
Adopting the extinction from the Balmer decrement (but cf.\ below) this corresponds 
to massive star populations of $\sim$ 25000 WNL, $\sim$ 10000--15000 WC, and 
$\sim$ 65000 equivalent O7V stars in an area of $\sim$ 1.7 $\times$ 3.5 kpc.
Several methods to estimate the metallicity concur to indicate \oh\ $\sim$
9.3--9.4, which makes Mkn 309 the most metal-rich WR galaxy known today.

The Mkn 309 observations have been combined with spectra of four other
metal-rich WR galaxies from the sample of Guseva \etal\ (2000)
to constrain the basic properties of the massive star populations
(IMF slope, \mup) and the star formation history (age, burst duration)
of these objects by detailed quantitative comparisons with appropriate 
evolutionary synthesis models (Sect.\ 5).
We have used the following main observational constraints:
H recombination lines, intensities and equivalent widths of the main WR features,
TiO bands originating from a population of late type stars (which are most likely
red supergiants; cf.\ Sect.\ 4)
and the detailed SED from $\sim$ 3500--7000 \AA.
Considering a large parameter space for the synthesis models
(metallicity, IMF slope and upper mass cut-off, star formation history, possible
effects of dust or photon leakage on the ionized gas, variable
stellar extinction) we obtain the following main results:

{\bf 1.)} The observations are explained by continuous star formation extending 
over $\sim$ 4--10 Myr (``extended bursts'') seen at ages of 7--15 Myr
or a superposition of several bursts with similar age differences 
(4--10 Myr) including a young ($\la$ 5 Myr burst).
This naturally explains both the observed WR populations (including WN and
WC stars) and the presence of red supergiants. 
Such extended burst durations, compared to other WR galaxies indicating
even shorter SF timescales (e.g.\ SCK99), are plausible in view of the
physical sizes of the observed regions and the nature and morphology of 
our objects (nuclear starbursts), and pose no fundamental physical problem
(Sect.\ 6).

{\bf 2.)} The SEDs in the optical range are very well reproduced for all 
objects, provided the stellar light suffers from a smaller extinction than
that of the gas (derived from the Balmer decrement).
Differences of $\Delta E(B-V) = $ 0.15--0.36 are found for our objects
with a mean $A_V=1.3 \pm 0.1$. 
Our finding of smaller stellar extinction based solely on optical data is 
in agreement with earlier studies (e.g.\ Fanelli \etal\ 1988, Calzetti 1997, 
Mas-Hesse \& Kunth 1999) requiring, however, the combination of UV and
optical data.

{\bf 3.)} The observational constraints are compatible with a Salpeter IMF
extending to masses \mup\ $\ga$ 40 \msun. Analysis of younger regions are 
required to provide real upper limit on the value of the upper mass cut-off.
We have also explored what deviations from this ``universal'' IMF are allowed.
Adopting a conservative approach we derive a {\em lower limit} of 
\mup $\sim$ 30 \msun\ for the Salpeter IMF.
Larger values (\mup $\ga$ 30--40 \msun) are obtained using probably more realistic 
assumptions on the metallicity and SF history.
An upper mass cut-off of at least $\sim$  35 -- 50 \msun\ seems also indicated 
from \hb\ equivalent width measurements of metal-rich \hii\ regions
in spiral galaxies (see Sect.\ 6).
Quite independently of the value of \mup, steeper IMFs with slopes such as the 
Miller \& Scalo (1979) IMF, are very unlikely.

Obviously the solutions fitting the observable constraints are not unique.
The main conclusions just summarised are, however, quite robust to changes
of the considered model parameters (see Sect.\ \ref{s_alt}).
Nevertheless it should be recalled that the results presented here rely
quite extensively on our present knowledge of massive stars, especially on 
the evolution of WR stars.
The advantage of studying objects of solar or
somewhat higher metallicity is that this corresponds to the domain with
the largest number of individual WR stars in the Local Group which has
previously been compared quite successfully to the evolutionary models used 
here (see Maeder \& Meynet 1994). In this sense the application of our 
synthesis models to starburst galaxies should at least allow meaningful
differential comparisons of properties such as the IMF.

Through the analysis of the massive star content 
in luminous metal-rich starbursts
by means of their {\em direct stellar features}, our aim is to 
constrain the upper end of the IMF (slope and upper mass cut-off) in metal-rich
environments in general and in particular in massive starbursts such as
ULIRGs. 
Whereas previous studies based on indirect indications from the ionized gas 
seem to indicate an IMF deficient in massive stars
(e.g.\ Goldader \etal\ 1997, Luhman \etal\ 1998, Bresolin \etal\ 1999,
but see the recent modeling efforts of Thornley \etal\ 2000),
our first results do not support such important differences compared to the
``standard'' Salpeter IMF.

To obtain more reliable constraints on the IMF in metal-rich environments
new high quality spectral observations will be necessary.
While analysis of few individual objects are extremely useful, larger
samples are e.g.\ required to constrain the exact value of \mup\
by sampling all relevant ages.  
To include the combined information from gas and stars, 
coupled evolutionary synthesis and photoionization models including all the 
stellar and nebular observables should be used consistently.
Furthermore the statistical distributions of 
the properties of e.g.\ metal-rich disk \hii\ regions also contain
encoded information about the IMF, SF history etc.
The theoretical tools for such analysis are available today.
With the new powerful telescopes available now, great progress should
be made on studies of stellar populations and the IMF in metal-rich
environments.

\begin{acknowledgements} 
We thank Richard Green for the participation
in the observations of Mkn 309 and valuable comments,
Liese van Zee, who kindly supplied us with some spectra of
metal-rich \hii\ regions, Paul Crowther for discussions on the 
classification of WR stars, and Miguel Cervi\~no for calculations 
and discussions on synthesis models including binary stars.
Claus Leitherer and Grazyna Stasi\'nska provided useful comments on an 
earlier version of this work. 
Yuri Izotov acknowledges support from the Observatoire Midi-Pyr\'en\'ees
where part of this work was done. 
DS, NGG and YII acknowledge the partial support of INTAS grant 97-0033.
TXT and YII acknowledge the partial financial 
support of NSF grant AST-9616863. He thanks the hospitality of 
the Observatoire of Paris-Meudon and the Institut d'Astrophysique in Paris.
\end{acknowledgements} 

{}

\clearpage

\begin{thebibliography}{999}
\bibitem[]{AHM88} Armus, L., Heckman, T. M., Miley, G. K., 1988, \apj\ 326, L45
\bibitem[]{AHM89} Armus, L., Heckman, T. M., Miley, G. K., 1989, \apj\ 347, 727
\bibitem[]{AKS89} Arnault, P., Kunth, D., Schild, H., 1989, \aap\ 224, 73
\bibitem[]{BM96} Bernasconi, P. A., Maeder, A., 1996, \aap\ 307, 829
\bibitem[]{BA86} Bica, E., Alloin, D., 1986, \aap\ 162, 21
\bibitem[]{B97} Bresolin, F., 1997, PhD thesis, University of Arizona
\bibitem[]{BKG99} Bresolin, F. , Kennicutt, R. C. , Jr., Garnett, D. R., 1999, \apj\ 
510, 104 
\bibitem[]{BH82} Burstein, D., Heiles, C., 1982, \aj\ 87, 1165
\bibitem[]{C97} Calzetti, D., 1997, in ``The Ultraviolet Universe at Low and High Redshift: 
  Probing the Progress of Galaxy Evolution'', Eds. W.H. Waller et al., AIP
                     Conference Proceedings, v.408., 403
\bibitem[]{C98} Cervi\~no, M., 1998, PhD thesis, Universidad Complutense, Madrid
\bibitem[]{CM94} Cervi\~no, M., Mas-Hesse, J.M.,  1994, \aap\ 284, 749 
\bibitem[]{} Cervi\~no, M., Mas-Hesse, J.M., Kunth, D., 2000, \aap, submitted
\bibitem[]{C91} Conti, P. S., 1991, \apj\ 377, 115
\bibitem[]{C99} Coziol, R., Reyes, R.E.C. Consid\`ere, S., Davoust, E., Contini, T.,
  1999, \aap\ 345, 733
\bibitem[]{CMB98} Crowther, P. A., de Marco, O., Barlow, M. J., 1998, \mnras\ 
296, 367
\bibitem[]{DR82} D'Odorico, S., Rosa, M., 1982, in ``Wolf-Rayet Stars: Observations,
  Physics, Evolution'', Eds. C. W. H. de Loore, A. J. Willis, IAU Symp.\ 99, 557
\bibitem[]{EP84} Edmunds, M. G., Pagel, B. E. J., 1984, \mnras\ 211, 507
\bibitem[]{FCT88} Fanelli, M. N., O'Connell, R. W., Thuan, T. X., 1988, \apj\ 
334, 665 
\bibitem[]{FN98} Figer, D. F., Najarro, F., 
  Morris, M. , McLean, I. S., Geballe, T. R., Ghez, A. M., Langer, N.,  
  1998, \apj\ 506, 384 
\bibitem[]{} Garc\'{\i}a-Vargas, M.L., Bressan, A., D\'{\i}az, A.I., 1995, \aaps\ 112, 13
\bibitem[]{} Garc\'{\i}a-Vargas, M.L., Gonz\'alez Delgado, R. M., P\'erez, E., Alloin, D.,
  D\'{\i}az, A.I., Terlevich, E., 1997, \apj\ 478, 112
\bibitem[]{G92} Garnett, D. R., 1992, \aj\ 103, 1330
\bibitem[]{GH98}  Gilmore, G., Howell, D., Eds., 1998, ``The Stellar Initial Mass Function'', 
  ASP Conf.\ Series, 142
\bibitem[]{GJ97} 
  Goldader, J. D., Joseph, R. D., Doyon, R., Sanders, D. B., 1997, \apj\  
  474, 104 
\bibitem[]{GLH99} Gonz\'alez Delgado, R. M., Leitherer, C., Heckman, T., 1999,
  \apjs\ 125, 489
\bibitem[]{GIT98} Guseva, N. G., Izotov, Y. I., Thuan, T. X., 1998, Kinematics and 
   Physics of Celestial Bodies 14, 1
\bibitem[]{GIT00} Guseva, N. G., Izotov, Y. I., Thuan, T. X., 2000, \apj\ 531, 776 (GIT00)
\bibitem[]{H97} Hamann, F., 1997, \apjs\ 109, 279
\bibitem[]{HG97} Heckman, T. M., 
  Gonzalez-Delgado, R., Leitherer, C., Meurer, G. R., Krolik, J., Wilson, A. 
  S., Koratkar, A., Kinney, A., 1997, \apj\ 482, 114 
\bibitem[]{HW99} Henry, R. B. C., Worthey, G., 1999, PASP 111, 919
\bibitem[]{HG99} Huang, J. H., Gu, Q. S., 
  Ji, L., Li, W. D., Wei, J. Y., Zheng, W., 1999, \apj\ 513, 215 
\bibitem[]{ITL94} Izotov Y. I., Thuan T. X., Lipovetsky V. A., 1994,
         ApJ 435, 647
\bibitem[]{ITL97} Izotov Y. I., Thuan T. X., Lipovetsky V. A., 1997,
         ApJS 108, 1
\bibitem[]{KG96} Kennicutt, R. C., Jr., Garnett, D. R., 1996, \apj\ 456, 504 
\bibitem[]{KJ99} Kobulnicky, H. A., Johnson, K .E., 1999, \apj\ 527, 154
\bibitem[]{KJ85} Kunth, D., Joubert, M., 1985, \aap, 142, 411
\bibitem[]{L98} Larson, R., 1998, \mnras\ 301, 569 
\bibitem[]{LSG99} Leitherer, C., Schaerer, D., Goldader, J. D., Gonz\'alez Delgado, R. M.,
  Robert, C., Foo Kune, D.,  De Mello, D., Devost, D., Heckman, T. M., 1999, \apjs\ 123, 3 
\bibitem[]{LSF98} Luhman, M. L., Satyapal, S., Fischer, J., et al., 
  1998, \apjl\ 504, L11 
\bibitem[]{MM94} Maeder, A., Meynet, G., 1994, \aap\ 287, 803 
\bibitem[]{MK99} Mas-Hesse, J. M., Kunth, D., 1999, \aap\ 349, 765 
\bibitem[]{MJ98} Massey, P., Johnson, O., 1998, \apj\ 505, 793 
\bibitem[]{MK97} McCarthy, J. K., 
  Kudritzki, R. -P. , Lennon, D. J., Venn, K. A., Puls, J.,  1997, \apj\  
  482, 757
\bibitem[]{M94} McGaugh, S. S., 1994, \apj\ 426, 135 
\bibitem[]{M95} Meynet, G., 1995, \aap\ 298, 767 
\bibitem[]{MMS94} Meynet, G., Maeder, A., Schaller, G., Schaerer, D., 
Charbonnel, C., 1994, \aaps\ 103, 97
\bibitem[]{MS79} Miller, G. E., Scalo, J. M., 1979, \apjs\ 41, 513
\bibitem[]{MH97} 
  Monteverde, M. I., Herrero, A., Lennon, D. J., Kudritzki, R. -P., 1997, 
  \apjl\ 474, L107 
\bibitem[]{N97} Najarro, F., Krabbe, A., 
  Genzel, R., Lutz, D., Kudritzki, R. P., Hillier, D. J., 1997, \aap\ 325, 
  700
\bibitem[]{O93} Oey, M. S., Kennicutt, R. C., Jr., 1993, \apj\ 411, 137 
\bibitem[]{OTT97} Ohyama, Y., Taniguchi, Y., Terlevich, R., 1997, \apjl\ 
480, L9 
\bibitem[]{O95} Olofsson, K., 1995, \aaps\ 111, 57 
\bibitem[]{OG99} Origlia, L., 
  Goldader, J. D., Leitherer, C., Schaerer, D., Oliva, E., 
  1999, \apj\ 514, 96 
\bibitem[]{OC82}  Osterbrock, D. E., Cohen, R. D., 1982, \apj\ 261, 64 (OC82)
\bibitem[]{} Pastoriza, M.G., Dottori, H.A., Terlevich, E., Terlevich, R.,
  Diaz, A.I., 1993, \mnras, 260 177 
\bibitem[]{} Raimann, D., Storchi-Bergmann, T., Bica, E., Melnick, J.,
  Schmitt, H., 2000, \mnras, in press (astro-ph/0004160)
\bibitem[]{S98} Scalo, J., 1998, in ``The Stellar Initial Mass Function'', 
  Eds. G. Gilmore, D. Howell, ASP Conf.\ Series, 142, 201
\bibitem[]{S96} Schaerer, D., 1996, \apjl\ 467, L17
\bibitem[]{S99a} Schaerer, D., 1999a, 
  in ``Wolf-Rayet Phenomena in Massive Stars and Starburst Galaxies'',
  IAU Symp.\ 193, 539 
\bibitem[]{S99b} Schaerer, D., 1999b, 
  in ``SpectroPhotometric Dating of Stars and Galaxies'', eds. I. Hubeny, S. Heap, 
  R. Cornett, ASP Conf. Series, 192, 49
\bibitem[]{SCK99} Schaerer, 
  D., Contini, T., Kunth, D., 1999a, \aap\ 341, 399 (SCK99)
\bibitem[]{SCP99} Schaerer, 
  D., Contini, T., Pindao, M., 1999b, \aaps\ 136, 35 
\bibitem[]{SV98} Schaerer, D., Vacca, W. D., 1998, \apj\ 497, 618 (SV98)
\bibitem[]{SBB99} Schiavon, R. P., Barbuy, B., Bruzual, G., 2000, \apj, 532, 453 
\bibitem[]{SFD98} Schlegel, D. J., Finkbeiner, D. P., Davis, M., 1998, \apj\ 500, 525
\bibitem[]{SV99} Schmutz, W., Vacca, W. D., 1999, New Astronomy 4, 197 
\bibitem[]{SS91} Shields, G. A., Skillman, E. D., Kennicutt, R. C., Jr., 1991, \apj\ 371, 82 
\bibitem[]{SC92} Silva, D. R., Cornell, M. E., 1992, \apjs\ 81, 865
\bibitem[]{S91} Smith, L. F. 1991, in ``Wolf-Rayet Stars and Interrelations with Other
        Stars in Galaxies'', IAU Symp.~143, eds. K. A. van der Hucht \& B. Hidayat,
        (Dordrecht: Kluwer), p.~601
\bibitem[]{SSM96} Smith, L. F., 
  Shara, M. M., Moffat, A. F. J., 1996, \mnras\ 281, 163 
\bibitem[]{St90} Stasi\'nska, G., 1990, \aaps\ 83, 501
\bibitem[]{St98} Stasi\'nska, G., 1998, in ``Dwarf Galaxies and Cosmology'', Eds. T.X. Thuan, 
  C. Balkowski, V. Cayette, J. Tran Thanh Van, Editions Frontieres (Gif-sur-Yvette, France), p. 259
\bibitem[]{SL96} Stasi\'nska, G., Leitherer, C., 1996, \apjs\ 107, 661
\bibitem[]{SS98} Surace, J. A., Sanders, 
  D. B., Vacca, W. D., Veilleux, S., Mazzarella, J. M., 1998, \apj\ 492, 116 
\bibitem[]{T90} Terlevich, E., Diaz, A., Pastoriza, M.G., Terlevich, R., Dottori, H., 1990, \mnras\ 242, 48P
\bibitem[]{T96} Terlevich, E., Diaz, A., Terlevich, R., Gonzalez-Delgado, R., Perez, E., Garcia-Vargas, M.L.,
  1996, \mnras\ 279, 1234
\bibitem[]{T00} Thornley, M.D., F\"orster-Schreiber, N.M., Lutz, D., Genzel, R., Spoon, H.W.W., 
  Kunze, D., 2000, \apj, in press (astro-ph/0003334)
\bibitem[]{VC92} Vacca, W. D., Conti, P. S., 1992, \apj\ 401, 543 
\bibitem[]{VK98} van der Hucht, K. A., Koenigsberger, G., Eenens, P. R. J., (Eds.), 
  ``Wolf-Rayet Phenomena in Massive Stars and Starburst Galaxies'', IAU Symp.\ 193
\bibitem[]{VSH98} van Zee, L., Salzer, 
  J. J., Haynes, M. P., O'Donoghue, A. A., Balonek, T. J., 
  1998, \aj\ 116, 2805 
\bibitem[]{VKS99} Veilleux, S., Kim, D.-C., Sanders, D. B., 1999, \apj\ 522, 113
\bibitem[]{W58} Whitford, A. E., 1958, \aj\ 63, 201
\end{thebibliography}
\end{document}